\documentclass[journal]{IEEEtran}
\ifCLASSINFOpdf
  \usepackage{graphicx}
  \usepackage{graphicx,charter, latexsym}

\newtheorem{lemma}{Lemma}
\usepackage{amssymb, amsmath,graphicx,charter, latexsym}
\usepackage{gensymb}
\usepackage{tabularx}

\newtheorem{theorem}{Theorem}

\newtheorem{example}{Example}
\usepackage{multirow,url,float,epstopdf,cite}
\usepackage{verbatim}

\usepackage{graphicx}
\usepackage{wrapfig} 
\usepackage{caption}
\usepackage{subcaption}
\usepackage{algorithmic}
\usepackage{algorithm}
\usepackage{tikz}
\usepackage{pifont}
\usepackage{xcolor}
\usepackage{caption}
\usetikzlibrary{patterns,calc}
\usetikzlibrary{shapes.geometric, arrows}
\usepackage[overlay,absolute]{textpos}

\newcommand\Max{\mathop{\mathcode`m=\mathcode`M \max}\displaylimits}

\tikzstyle{block}=[draw opacity=0.7,line width=1.4cm]
\tikzstyle{ag} = [circle, radius =3cm, text centered, draw=black]
\tikzstyle{startstop} = [rectangle, rounded corners, minimum width=3cm, minimum height=1cm,text centered, draw=black, fill=red!30]
\tikzstyle{io} = [trapezium, trapezium left angle=70, trapezium right angle=110, minimum width=1cm, minimum height=1cm, text centered, draw=black, fill=blue!30]
\tikzstyle{process} = [rectangle, draw,fill=orange!30, text width = 20em, text centered, rounded corners, minimum height=4em, minimum width=1cm]
\tikzstyle{upd} = [rectangle, draw,fill=orange!30, text width = 5em, text centered, rounded corners, minimum height=4em, minimum width=1cm]
\tikzstyle{decision} = [diamond, minimum width=3cm, minimum height=1cm, text centered, draw=black, fill=green!30]
\tikzstyle{arrow} = [thick,->,>=stealth]
\tikzset{
  solid node/.style={circle,draw,inner sep=1.2,fill=black},
  hollow node/.style={circle,draw,inner sep=1.2},
}

\DeclareTextFontCommand{\emph}{\em}

\IEEEoverridecommandlockouts

\else
\fi
\begin{document}
\title{Throughput Optimal Decentralized Scheduling of Multi-Hop Networks with End-to-End  Deadline Constraints: Unreliable Links}

\author{Rahul Singh~\IEEEmembership{Member,~IEEE,} and
        P.~R.~Kumar~\IEEEmembership{Fellow,~IEEE,}
        % <-this % stops a space
\thanks{R. Singh is at 32-D716, LIDS, MIT, Cambridge, MA 02139; P. R. Kumar is
at 
Dept. of ECE, Texas A\&M Univ., 3259 TAMU, College Station, TX 77843-3259).
{\tt\small rsingh12@mit.edu, prk@tamu.edu.}
}% <-this % stops a space
\thanks{Preferred address for correspondence: P. R. Kumar, Dept. of ECE, Texas A\&M Univ., 3259 TAMU, College Station, TX 77843-3259.}% <-this % stops a space
\thanks{This material is based upon work partially supported by NSF under Contract
Nos. CNS-1302182 and
NSF Science \& Technology Center Grant CCF-0939370, and
USARO under Contract W911NF-15-1-0279.}}


\markboth{}%
{Shell \MakeLowercase{\textit{et al.}}: Throughput Optimal Decentralized Scheduling of Multi-Hop Networks with End-to-End  Deadline Constraints: Unreliable Links}



\maketitle

\begin{abstract}
We consider multi-hop networks serving multiple flows in which packets not delivered to their destination nodes by their deadlines are dropped from the network. The throughput of packets that are delivered within their end-to-end deadlines is called the timely-throughput. We address the design of policies for routing and scheduling packets that optimize any specified weighted average of the timely-throughputs of several flows, under
nodal power constraints. 

We provide a new approach which directly yields an optimal distributed scheduling policy that attains any desired maximal timely-throughput vector (i.e., any point on the Pareto frontier) under average-power constraints on the nodes. 
This completely distributed and tractable solution structure arises from a novel intrinsically stochastic decomposition of the Lagrangian of the constrained network-wide
Markov Decision Process rather than of the fluid model.

The derived policies are highly decentralized in several ways. All decisions regarding a packet's transmission scheduling, transmit power level, and routing, are based solely on the age of the packet, not requiring any knowledge of network  state or queue lengths at any of the nodes. Global coordination is achieved through a ``price" for energy usage paid by a packet each time that its transmission is attempted at a node. This price decouples packets entirely from each other. It is different from that used to derive the backpressure policy where price corresponds to queue lengths.
%, and nodes need to share knowledge of their queue lengths with neighbors. 

The complexity of calculating the prices is tractable, being related only to the number of nodes multiplied by the relative deadline bound, and is considerably smaller than the number of network states which is exponentially large. Prices can be determined offline and stored. 

If nodes have peak-power constraints instead of average-power constraints, then the decentralized policy obtained by truncation is near-optimal with respect to the timely-throughput as link capacities increase in a proportional way.
%, in the same quantifiable sense as Whittle's relaxation is near-optimal as the numbers of bandits of different types increase in a proportional way.
\end{abstract}

\begin{IEEEkeywords}
Communication Networks, Wireless Networks, Delay Guarantees in Networks, Scheduling Networks, Quality of Service.
\end{IEEEkeywords}

% \fi
\IEEEpeerreviewmaketitle

%\section{Introduction}
%\IEEEPARstart{T}{his} demo file is intended to serve as a ``starter file''
%for IEEE journal papers produced under \LaTeX\ using
%IEEEtran.cls version 1.8b and later.
%% You must have at least 2 lines in the paragraph with the drop letter
%% (should never be an issue)
%I wish you the best of success.

\section{Introduction}\label{Introduction}
%Applications such as cyber-physical systems, where control-loops are closed over networks, are sensitive to delays. Similarly, Quality of Service (QoS) requirements for real-time applications such as video streaming, VoIP, surveillance, sensor networks, mobile ad-hoc networks (MANETS), and in-vehicular networks, entail that packets should be delivered on time~\cite{jandrews}. The above objective may also have to be achieved in an energy efficient manner. Further, due to the distributed nature of the multi-hop network, the scheduling of packets by nodes will need to be done in a decentralized manner.

\IEEEPARstart{T}he past quarter century has seen the pioneering work of Tassiulas and Ephremides \cite{tassi1},
Lin and Shroff \cite{shroffcdc},
Lin, Shroff and Srikant \cite{linshrof}, 
and Neely, Modiano and Rohrs \cite{modiano1} on Max-Weight and backpressure based
scheduling policies for communication networks that are provably throughput optimal, attaining any desired maximal throughput vector on the Pareto frontier of the feasible throughput region. In this paper we  develop completely decentralized and tractable scheduling policies that achieve any desired maximal throughput vector of packets that meet specified hard end-to-end relative deadlines for packets under average-power constraints on nodes.

\begin{figure}
\vspace{-0.25in}
\includegraphics[trim = 50mm 65mm 20mm 20mm, clip, width=10cm]{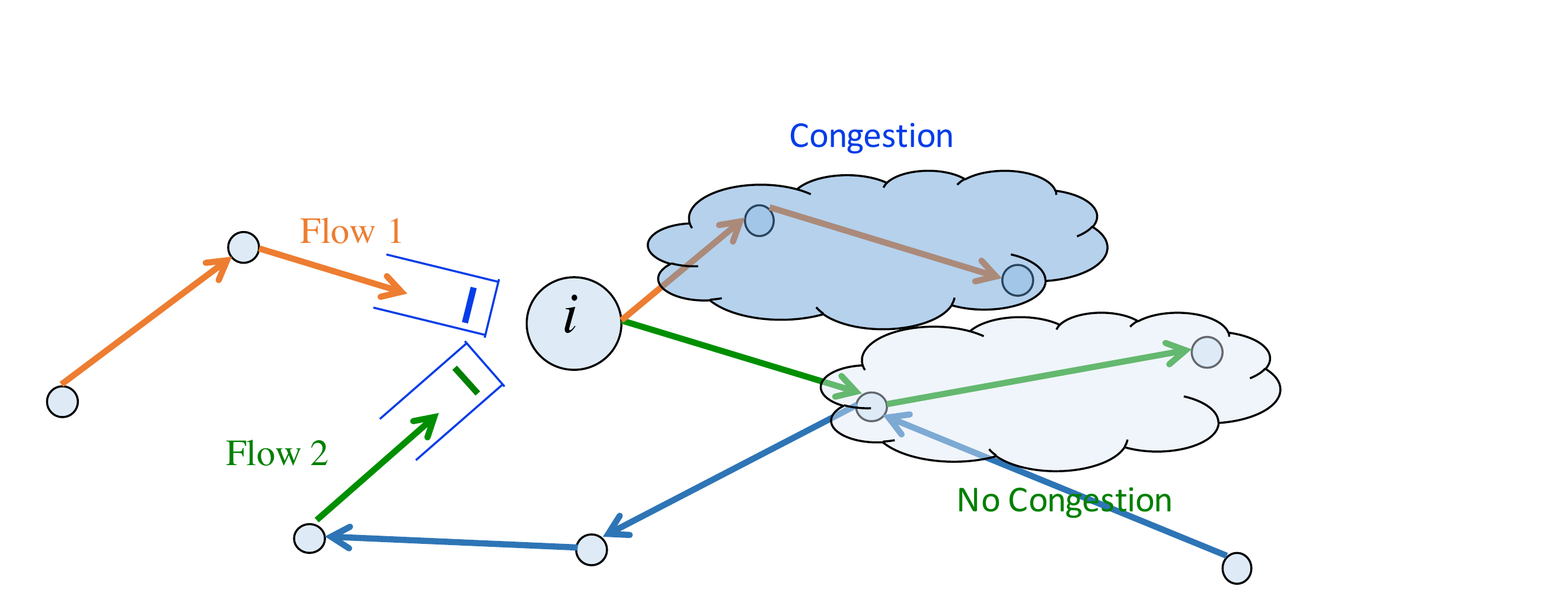}
\caption{The challenge of distributed scheduling. Suppose that Node $i$ wishes to transmit one packet,
either a packet of Flow $1$ or a packet of Flow $2$. Suppose packets of Flow $1$ are experiencing downstream congestion, but packets of Flow $2$ face no downstream congestion. Then serving a packet of Flow 1 is not useful since it will just get stuck in downstream congestion and not make it to its destination in time.
Hence Node $i$ should serve a packet of Flow $2$. Therefore network state is useful information. 
%Thus optimal scheduling under end-to-end deadlines benefits from network state information at all nodes.
This creates a chicken and egg situation since obtaining instantaneous network state information requires zero-delay communication of information over the network, while the very purpose of obtaining network state information is to provide low delay communication. We show that this conundrum can be completely resolved when nodes
have average-power constraints.}
\label{fig2}
\end{figure}

To see why this is a challenge, one may consider the situation depicted in Figure~\ref{fig2}. Suppose that Node $i$ needs to decide whether to serve the packet of Flow $1$ or the packet of Flow $2$. Packets of Flow $1$ are experiencing downstream congestion, in contrast to packets of Flow $2$ which face no downstream congestion. Thus Node $i$ is better off serving the packet of Flow $2$ rather than Flow $1$ since the latter would anyway get delayed and not make it to its destination on time. Therefore network state is useful information. The central contribution of this paper
shows just how to obtain an optimal distributed scheduling
policy when nodes face average-power constraints. When nodes face peak-power constraints we provide a distributed approximately optimal policy that approaches optimality in a precisely quantifiable way.
%\begin{figure}
%\vspace{-0.5in}
%\centering
%\includegraphics[width=2.0\linewidth]{figures/Fig-1.pdf}
%\vspace{-0.65in}
%\caption{System considered in Example~\ref{ex2}.}
%\label{fig4}
%\end{figure}

The main contribution is the design of completely decentralized optimal
routing and scheduling policies which can attain any desired maximal throughput vector of packets that are delivered end-to-end by their stipulated deadlines, when nodes have average-power constraints. That is, the policies are precisely optimal with respect to the throughput 
delivered under hard end-to-end delay constraints. 
These results addressing per-packet hard delay bounds are obtained by considering a decomposition of the
Lagrangian of the constrained network-wide Markov Decision Process that is intrinsically stochastic and different from a fluid-based analysis.
We show that a policy
where each node makes decisions based only on the age of the packets present at it, and a prior computable price of transmission, 
oblivious to all else in the network, is optimal. This vastly simplifies the network operation.
% compared to policies in which a node requires the knowledge of queue lengths at other nodes. 
 If the nodes instead have peak-power constraints, then the decentralized policy can be simply truncated to yield a policy that is near-optimal in the same quantifiable sense as Whittle's relaxation for multiarmed bandits \cite{whittle,weiss}.

In this paper we address the case where links are unreliable. 
This is of interest in networks with directional antennas, or networks of microwave repeaters, or even 
networks composed of unreliable wireline links. 
In a companion paper \cite{sinkum16b}  we address the case where the links face interference.

Delivering packets on time is of great interest in emerging applications such as cyber-physical systems, where control-loops are closed over networks, are sensitive to delays. Similarly, Quality of Service (QoS) requirements for real-time applications such as video streaming, VoIP, surveillance, sensor networks, mobile ad-hoc networks (MANETS), and in-vehicular networks, all entail that packets should be delivered on time~\cite{jandrews}. 

The rest of this paper is organized as follows.
In Section \ref{Problem-Considered} we describe the problem considered and the main results.
In Section \ref{pw} we summarize previous work and set this work in context.
In Section \ref{sm} we describe the system model.
In Section \ref{rr} we describe the constrained network-wide MDP.
In Section \ref{a1} we show the packet-level decomposition of the network-wide Lagrangian.
In Section \ref{spsf} we describe the Single-Packet Transportation MDP that arises.
In Section \ref{pmdp} we consider the Dual Problem of the constrained network-wide MDP and establish strong duality.
In Section \ref{solution} we show that the implementation of the packet-level optimal transportation policies yields
an overall optimal policy for the entire system.
In Section~\ref{direct-sol} we show how the overall optimal policy is obtained through
a tractable linear program.
In Section \ref{threshold} we show that there is a further simplifying threshold structure for the optimal policy.
In Section \ref{sec:optpol} we show how the optimal prices can be precomputed.
In Section \ref{sec:linkcap} we address the problem with link capacities or, equivalently, peak-power constraints.
In Section \ref{sec:peakpow} we establish the asymptotic optimality of the truncated policy.
In Section \ref{wirelessfading} we address the problem when the channel condition changes with time.
In Section \ref{jointnonreal} we address the problem where there are both real-time flows as well as non-real-time flows.
In Section \ref{examples} we provide examples showing how the theory can be used to determine optimal distributed policies,
and also present a comparative simulation study of the performance of the truncated policy for link-capacity constraints.
We conclude in Section \ref{conclusion}.

\section{Problem Considered and Main Results} \label{Problem-Considered}
We consider multi-hop, multi-flow networks in which packets of all flows have a
hard end-to-end relative deadlines.
(The relative deadline is the remaining time-till-deadline when a packet arrives).
%The results extend trivially to situations where the specification of the relative deadline is allowed to vary from packet to packet. 
Nodes can transmit packets at varying power levels. Since the wireless channel is unreliable, the outcome of packet transmissions is modeled as a random process. Nodes can transmit and receive packets simultaneously.  The throughput of packets of a flow that meet the end-to-end relative
deadline constraint is called the timely-throughput. We consider the following two types of nodal constraints in this paper: a)  An average-power constraint on each node in the network, or b) A link-capacity constraint on each network link which bounds the number of concurrent packets that can be transmitted on it at any given time $t$, or,
equivalently, a peak-power constraint at each node.  

Our goal is to design a decentralized, joint scheduling, transmission and routing policy, abbreviated
as a ``scheduling policy," that maximizes the weighted sum of the timely-throughputs of the flows, for any given nonnegative choice of weights.
That is, it can attain any point on the Pareto frontier of the timely-throughput vector.
To be optimal, the policy should be dynamic enough and take into account in an online fashion the following factors:
%while taking decisions in order to serve our purpose:
\begin{itemize}
\item \textit{Routing}: The policy will need to dynamically route packets so as to avoid paths that have a higher delay or nodes
with lower power budgets. 
 \item \textit{Scheduling}: The policy will need to prioritize packet transmissions based on their age, the channel conditions, and the congestion at the nodes lying on the paths to their destinations.
\item \textit{Energy Efficiency and Channel Reliability}: 
It will need to choose the power levels of packet transmissions to balance between reliability and energy consumption. If channel states are time-varying, the policy will have to carry out packet transmissions opportunistically when the channel states are ``good" so that the maximum throughput under deadline constraints is attained in an energy-efficient manner~\cite{modiano1,berry,salodkar}. This also involves a trade-off between packets missing their deadlines on account of bad channel conditions, and between spending more energy to transmit it in a bad channel state so that it reaches the destination within its deadline.
\end{itemize}

The main result is the determination that a 
Markov Decision Process 
for a certain ``Single-Packet Transportation Problem" governs the behavior of each packet, oblivious to all other traffic or network state.
In this standalone problem, a single packet optimizes its progress through the network,
paying prices to nodes every time it requests transmission,
but is compensated with a reward if it reaches its destination prior
to the hard deadline. 
The only manner in which this optimal single-packet transportation problem is coupled to the overall network,
nodal power constraints, other flows and 
other packets, is through predetermined prices for nodal transmissions.
The optimal prices can be tractably computed off-line and stored.
Determining the optimal policies for all packets is also of tractable complexity,
involving a linear program with the number of variables equal to the 
product of the square of
the number of nodes, the number of flows and the maximum relative deadline,
rather than the network state that is exponential in the above quantity. 
Thereby, we obtain optimal distributed policies for maximizing the network's timely-throughput of packets meeting hard
per-packet deadlines under average-power constraints on the nodes.
%If nodes have peak-power constraints, then the above policy can simply be truncated to obtain a near-optimal policy.

The key to these results is to pursue a fundamentally stochastic approach
that considers the Lagrangian of the constrained network-wide Markov Decision Process (MDP) governing the
entire network, and showing how it decomposes into packet-by-packet decisions.
This decomposition approach allows treatment of the intrinsically variability
related aspects such as delay, in sharp contrast to the backpressure approach that considers the decomposition of the Lagrangian of the
fluid model.
Thus the approach of this paper is able to address delay rather than just throughput.
Through this novel decomposition, we can address timely-throughput optimality of
packets that meet hard per-packet delay deadlines, rather than just throughput optimality.
Moreover, it does so
producing a completely distributed and tractable policy, for systems with average-power constraints at nodes and unreliable links.

When nodes have peak-power constraints in addition to, or in place of, average-power constraints, one can simply truncate the above optimal policy dynamically to respect the peak-power constraint,
and obtain a policy that is quantifiably
near-optimal. Specifically, it is asymptotically optimal in the same manner
as Whittle's relaxation for multi-armed bandits
\cite{whittle,weiss}.

%Our approach to derive an optimal 
%or near-optimal policy 
%is as follows. 
The exposition proceeds as follows.
We begin with the overall
problem with average nodal power constraints and invoke the scalarization principle to pose the problem of maximizing the network's weighted timely-throughput subject to nodal average-power constraints as a constrained network-wide Markov Decision Process (MDP)~\cite{altman}. We then solve this problem by considering the Lagrangian dual of this constrained network-wide
MDP. The Lagrange multipliers associated with the average-power constraints are interpreted as prices paid by a packet for utilizing energy every time its transmission is attempted by a node. 
As recompense, a packet collects a reward equal to the weight of its flow when it reaches its destination within its specified hard deadline. This results in a very
convenient packet-level decomposition into an Optimal Single-Packet Transportation Problem.
%This is solved by a Single-Packet Transportation Problem MDP. 
The Markov Decision Process for this problem
has a small-sized state-space: (Number of nodes)*(Bound on relative deadline), much smaller than the
exponentially large number of states in the network-wide problem.
The prices can be calculated off-line just by price tattonement that drives the ``excess power consumption" at nodes to zero. 
The optimal policies for all packets of all flows
can be determined by solving a linear program with a small number of states.
Importantly, the overall approach yields a completely distributed solution, where a node only needs to know the remaining times-till-deadlines of
packets present at that node, that is
timely-throughput optimal under average-power constraints at nodes.

When the constraints are on link-capacity and not its average utilized capacity, or equivalently
on peak-power and not average-power, one obtains a near-optimal policy by simply
truncating the average power-optimal policy. The result is asymptotically optimal as the network capacity is increased, in the same sense as Whittle's indexability~\cite{whittle} approach is asymptotically optimal as the population of bandits increases in proportion~\cite{weiss}.

\section{Previous Works}\label{pw}
Over the past twenty-five years there have been several notable advances in the
theory of networking.

In pioneering work, Tassiulas and Ephremides \cite{tassi1}, Lin and Shroff \cite{shroffcdc},
Lin, Shroff and Srikant \cite{linshrof}, and Neely, Modiano and Rohrs \cite{modiano1}
have shown that scheduling networks based
on Max-Weight and backpressure are throughput optimal.
The backpressure policy emerges naturally from a decomposition of the Lagrangian for the fluid network.

In another breakthrough, Jiang and Walrand \cite{wlrnd1} have designed a novel 
Adaptive Carrier
Sensing Multiple Access algorithm for a general interference
model that achieves maximal throughput through completely
distributed scheduling under slow adaptation, without slot synchrony,
if packet collisions are ignored.
Combined with end-to-end control it also achieves fairness among the multiple flows.

In another seminal contribution, Kelly, Maulloo and Tan \cite{kelly} have shown that the problem of congestion control of the Internet can be formulated as convex programming problem and have provided
a quantitative framework for design based on primal or dual approaches.

Eryilmaz, Ozdaglar and Modiano \cite{eryilmaz2007polynomial} have developed a 
throughput optimal randomized algorithm for routing and scheduling of
the common two-hop interference model that can be implemented in a distributed way
with polynomial complexity. They have also developed such a policy for inelastic
flows that takes flow control into account and results in a fair allocation of the network's capacity.

Any effort at developing a delay-optimal scheduling policy needs to take into account the time-till-deadline of packets in the network.
%, and the channel reliabilities of the links that the packets have to traverse in order to reach the destination node. 
The CSMA algorithm does not do so.
The backpressure policy schedules packets only on the basis of rate-weighted queue lengths of nodes and provides no delay guarantees. 
In fact, fluid-based policies such as  the backpressure policy, should be expected to and have been shown to be throughput optimal, but they
should not be expected to provide delay optimality.
They can perform poorly with regard to delay performance~\cite{d1,d2,d3,d4}. 
For optimal delay performance, one needs to start with a fundamentally stochastic framework that takes fluctuations into account. 
This is akin to the difference between the
law of large numbers and the central limit theorem.
Such a stochastic framework is the path that is pursued in this paper.
%This is one key reason why it can result in a high end-to-end delay. 
%Moreover the backpressure policy requires the nodes to share the information of their queue lengths with their neighboring nodes, and thus is not decentralized. 
%Reference~\cite{cdcdelay} considers the setting of multi-hop networks in which the routes for each flow are fixed, and devises a scheduling policy that is highly decentralized.

There has been considerable progress on the problem of scheduling an access point,
in which multiple one-hop flows with hard relative deadlines share a wireless channel.
The Pareto optimal frontier of timely-throughput vectors has been characterized, and simple optimal policies have been determined \cite{Hou2009,houkum09b,houkum10,IHHou10Utility,IHHou10Hoc,hou2011broadcasting,hou2011optimality,hou2011scheduling,hou2012real,hou2011survey,houqueueing,houkum13, rahul1,rahul}.
%In such a network, flows only traverse a single hop. 

Li and Eryilmaz~\cite{atilla3} consider the problem of scheduling deadline-constrained packets over a multi-hop network. However, the proposed policies are not shown to have any provable guarantees on the resulting  timely-throughput. To the best of the authors' knowledge, Mao, Koksal and Shroff~\cite{shroff1} is the only work which provides a provable sub-optimal policy for deadline-constrained networks, though the policies proposed therein guarantee only a fraction, 
${\mbox{(Length of the longest route in the network)}}^{-1}$, of the timely-throughput capacity region. 
%\begin{align*}\frac{1}{\mbox{length of the longest route in the network}},\end{align*} of the timely-throughput capacity region. 

%The rest of this paper is organized as follows. 
%We will first focus on the problem of joint transmission, routing and scheduling under the average-power constraints~\eqref{shannon}, and will then address the corresponding problem with link-capacity constraints~\eqref{linkcap} in Section~\ref{sec:linkcap}. 

\section{The System Model }\label{sm}
We consider networks in which the data-packets have a hard deadline constraint on the time within which they should be delivered to their destination nodes if they are to be counted in the throughput. 
%We will describe below the model and detail the assumptions under which the results will be established.
%We will also indicate extensions of the assumptions under which the results continue
%to hold, but for brevity and simplicity of exposition we will not carry through the extra notations that would
%be required in the general case.

\begin{figure}
\vspace{-0.75in}
\begin{center}
\includegraphics[width=3in, angle=270]{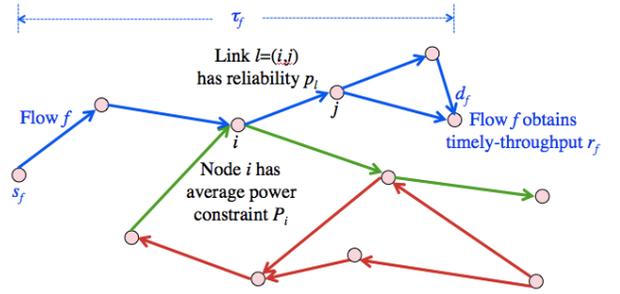}
\end{center}
\vspace{-0.75in}
\caption{Multi-hop network serving $F$ flows. Flow $f$, with source $s_f$ and destination $d_f$, has several feasible routes. Its end-to-end relative deadline is $\tau_f$. Node $i$ has an average-power constraint $P_i$. A packet transmitted on link $\ell = (i,j)$ has a probability $p_\ell$ of being successfully received by node $j$. Though not so indicated, the probability $p_i$ may depend on the power level of the packet's transmission by node $i$. 
%Link $\ell$ can transmit.
}
\label{fig1}
\end{figure}

The communication network of interest is described by a directed graph $G=\left(V,\mathcal{E}\right)$ as shown in Figure~\ref{fig1}, where $V = \{1,2,\ldots,|V|\}$ is the set of nodes that are connected via communication links. A directed edge $ (i,j) \in \mathcal{E}$ signifies that node $i$ can transmit data packets to node $j$. We will call this link $\ell =(i,j)$.
%\begin{comment}
%\begin{figure}[!t]
%	\centering
% \resizebox{7cm}{5cm}{
%\begin{tikzpicture}
%\node (so) at (-1,0) [draw,circle,minimum size=.25cm] {$s$};
%\node (a) at (0,0) [draw,circle,minimum size=.25cm] {$2$};
%\node (b) at (1,0) [draw,circle,minimum size=.25cm] {$5$};
%\node (c) at (2,0) [draw,circle,minimum size=.25cm] {$d$};
%\node (d) at (0,1) [draw,circle,minimum size=.25cm] {$1$};
%\node (e) at (0,-1) [draw,circle,minimum size=.25cm] {$3$};
%\node (f) at (1,1) [draw,circle,minimum size=.25cm] {$4$};
%\node (g) at (1,-1) [draw,circle,minimum size=.25cm] {$6$};
%
%\draw [->] (a) -- (b);
%\draw [->] (b) -- (c);
%\draw [->] (d) -- (a);
%\draw [->] (a) -- (e);
%\draw [->] (f) -- (b);
%\draw [->] (b) -- (g);
%\draw [->] (so)--(d);
%\draw [->] (so)--(a);
%\draw [->] (so)--(e);
%\draw [->] (d)--(f);
%\draw [->] (e)--(g);
%\draw [->] (f)--(c);
%\draw [->] (g)--(c);
%
%%\draw [->] (c) -- (a);
%\end{tikzpicture}}
%\caption{Multi-hop sensor network serving a single flow with source node $s$, and destination node $d$. Each directed edge corresponds to a link.}
%\label
%\end{figure}
%\end{comment}
% trim={<left> <lower> <right> <upper>}
%\begin{figure}
%\includegraphics[trim = 60mm 40mm 20mm 20mm, clip, width=10cm, width=0.5\linewidth, angle=270]{figures/Fig-0.pdf}
%\caption{Multi-hop network serving $F$ flows. Flow $f$, with source $s_f$ and destination $d_f$, has several feasible routes. Its end-to-end relative deadline is $\tau_f$. Node $i$ has an average-power constraint $P_i$. A packet transmitted on link $\ell = (i,j)$ has a probability $p_\ell$ of being successfully received by node $j$. Though not so indicated, the probability $p_i$ may depend on the power level $E$ of the packet's transmission by node $i$, and more generally even on the time $t$ that it is transmitted. 
%%Link $\ell$ can transmit.
%}
%\label{fig1}
%\end{figure}
%\begin{figure}
%\includegraphics[trim = 60mm 40mm 20mm 20mm, clip, width=10cm]{figures/Fig-0.pdf}
%\caption{Multi-hop network serving $F$ flows. Flow $f$, with source $s_f$ and destination $d_f$, has several feasible routes. Its end-to-end relative deadline is $\tau_f$. Node $i$ has an average-power constraint $P_i$. A packet transmitted on link $\ell = (i,j)$ has a probability $p_\ell$ of being successfully received by node $j$. Though not so indicated, the probability $p_i$ may depend on the power level $E$ of the packet's transmission by node $i$, and more generally even on the time $t$ that it is transmitted. 
%%Link $\ell$ can transmit.
%}
%\label{fig1}
%\end{figure}

We assume that time is discrete, and evolves over slots numbered $1,2,\ldots$. One time-slot is the time taken to attempt a packet transmission over any link in the network. 

There are a finite number of transmit power levels at which
a packet can be transmitted. For convenience, we normalize each time slot to 1 second, so that power and energy of a transmission are interchangeable.

The outcome of a transmission over a link between any two nodes is allowed to be random, which enables us to model unreliable channels.
If a packet transmission occurs on the link $\ell$ at a certain power level that consumes energy $E$, then the transmission is successful with probability $p_\ell(E)$, which is monotone decreasing in $E$. We can model the phenomena of wireless fading by allowing the success probability $p_\ell(t,E)$ to also be a function of time that can be assumed to be governed by a finite-state Markov process, whose state is known at the transmitting node. 
%We can also incorporate transmit power control by allowing the success probability $p_\ell(t,E)$, to depend on the transmission power $E$.  
However, for simplicity of exposition, we consider time-invariant $p_\ell(E)$'s only.
%, and suppose that $E \equiv 1$ for all transmissions. 
In this paper we do not consider contention for the transmission medium; the case of interference is considered in the companion paper \cite{sinkum16b}. 

The network is shared by $F$ flows. Packets of flow $f$ have source node $s_f$ and destination node $d_f$. They may traverse one of several alternative routes.

 The numbers of packet arrivals of a flow at its source node are i.i.d. across time-slots, though the distribution can vary from flow to flow. For simplicity of exposition we suppose that these distributions have bounded support, i.e., the number of arrivals is bounded, though we can relax this to merely assuming they are finite valued. Packets across flows are independent. The analysis below carries over to the case when the arrivals and relative deadlines (detailed below) are governed by a finite-state Markov process. We will denote the average arrival rate of flow $f$ in packets/time slot by $A_f$.

Each packet of flow $f$ has a ``relative-deadline," or ``allowable delay" $\tau_f$.
If a packet of flow $f$ arrives to the network
at time $t$, then it needs to be either delivered to its destination node by time-slot $t + \tau_f$ ,
or else it is discarded from the network at time $t + \tau_f$ if it has not yet reached its destination $d_f$. 
We suppose that all relative deadlines of packets are bounded by a quantity $\Delta$.
We can allow packets of a flow to have random independent and identically distributed (i.i.d.) deadlines, independent across flows, however, for simplicity of exposition we will suppose that all packets of flow $f$ have the same relative deadline $\tau_f$. 

The ``timely-throughput" $r_f$ attained by a flow $f$ under a policy is the average number of packets delivered prior to deadline expiry per unit time, i.e., 
%\begin{eqnarray}\label{tp}
%&\liminf_{T\to\infty} \frac{1}{T}\mathbb{E} \sum_{t=1}^{T}\delta_f(t),
%\end{eqnarray} 
\begin{align}\label{tp}
r_f := \liminf_{T\to\infty} \frac{1}{T}\mathbb{E} \sum_{t=1}^{T}\delta_f(t),
\end{align} 
where the random variable $\delta_f(t)$ is equal to the number of packets of a packet of flow $f$ that
are delivered in time to their destination at time $t$, with the expectation taken under the policy being applied. 

The vector $\boldsymbol{r} := (r_1, r_2, \ldots, r_F)$ is called the ``timely-throughput vector."
A timely-throughput vector $\boldsymbol{r}$ that can be achieved via some scheduling policy will be called an ``achievable timely-throughput vector". The set of all achievable timely-throughput vectors constitutes the ``rate-region," denoted by $\boldsymbol{\Lambda}$. 
%A scheduling policy that achieves the complete rate-region 
%$\boldsymbol{\Lambda}$ is said to be timely-throughput-optimal. Thus, under the application of a timely-throughput-optimal policy $\pi$, the network can achieve any timely-throughput vector that can be attained by some scheduling policy. 

Given weights $\beta_i \geq 0$ for the timely-throughput of each flow $f_i$, we will define the
``weighted timely-throughput" as $\beta^T \boldsymbol{r}$, where
$\beta = (\beta_1,\beta_2,\ldots,\beta_F)$.

%We will consider below a ``weighted timely-throughput," defined as $\beta^T \boldsymbol{r}$, with
%$\beta = (\beta_1,\beta_2,\ldots,\beta_F)$ where $\beta_i \geq 0$ denotes the weight given to the timely-throughput flow $f_i$.
%
In Sections \ref{rr}-\ref{sec:optpol} we consider an average-power constraint on each node $i\in V$. If the total energy consumed by all the concurrent packet transmissions on link $\ell$ at time $t$ is $E_\ell(t)$ units of energy, then the nodal average-power constraints are given by,
\begin{align}\label{shannon}
\limsup_{T\to\infty}\frac{1}{T}\mathbb{E} \sum_{t=1}^{T}\sum_{\ell:\ell=(i,\cdot)}E_\ell(t) \leq P_i, \forall i\in \{1,2,\ldots,V\}.
\end{align} 
The second summation above is taken over all links $\ell$, where $\ell = (i,j)$ for some node $j$.
For simplicity of exposition, we suppose that the number of power levels available to choose from for any transmission is finite. We note that the above constraint on the average-power allows a node to transmit packets simultaneously over several outgoing links, which can be achieved via employing various techniques such as TDMA, OFDMA, CDMA etc., \cite{ofdma1,ofdma2,ofdma3}. We suppose that nodes can simultaneously receive any number of packets while they are transmitting. 

We will derive completely distributed scheduling
policies that maximize the weighted timely-throughput for any given weight vector $\beta$ for the unreliable multi-hop network under the average-power constraint (\ref{shannon}) on nodes.

As an alternative to \eqref{shannon}, or in addition to it, in Section \ref{sec:linkcap} we will consider peak-power constraints on each link:
\begin{align}\label{linkcap}
E_\ell(t) \leq C_{\ell}, \forall \ell \in \mathcal{E},\mbox{ and } t=1,2,\ldots.
\end{align} 
Alternatively, we can constraint the number of concurrent packets that can be transmitted on a link $\ell$ at each time $t$. For 
either of these situations, we will obtain quantifiably near-optimal distributed scheduling policies.

%The focus of our analysis in Sections \ref{rr}-\ref{sec:linkcap}
%will be on the problem with nodal average-power constraints~\eqref{shannon}.
%It should be noted that the nodal average-power constraints then reduce to constraints on the average node usage when the transmission power is fixed, i.e., the average number of times a node schedules packet transmissions.

\section{Characterizing the Rate Region: The Constrained Network-Wide MDP}\label{rr}
In order to characterize the network's rate-region $\boldsymbol{\Lambda}$, it is sufficient to characterize the set of Pareto-optimal vectors
$\{\boldsymbol{r}: \boldsymbol{r}$ is a timely-throughput vector and $\exists \beta\in \mathbb{R}^{N}_{+}$  such that 
$\boldsymbol{r} \in \arg\max_{\boldsymbol{y}\in \Lambda} \sum_{f} \beta_f  y_f \}$,
%\begin{align*}
%&\{\boldsymbol{r}: \boldsymbol{r}\mbox{ is a timely-throughput vector and }\exists \beta\in \mathbb{R}^{N}_{+} \mbox{ such that } \\
%& \qquad \boldsymbol{r} \in \arg\max_{\boldsymbol{y}\in \Lambda} \sum_{f} \beta_f  y_f    \},
%\end{align*}
since $\boldsymbol{\Lambda}$ is simply its closed convex hull. The problem of obtaining $\boldsymbol{\Lambda}$ therefore reduces
to that of finding scheduling policies which maximize
weighted timely-throughputs of the form $\beta^T \boldsymbol{r}$.

%\subsection{Constrained MDP Formulation}
The latter problem can be posed as a Constrained Network-wide Markov Decision Process (CMDP)~\cite{altman1}. 
The state of an individual packet present in the network at time $t$ is described by the flow $f$ to which it belongs, and the two tuple $\left(i,s\right)$, where $i$ is the node at which it is present, and $s$ is the time-to-go till its deadline. The state of the network at time $t$, $X(t)$, is described by specifying the state of each packet present in the network at time $t$.
Since the time spent by a packet in the network is bounded by $\Delta$, and since the number of arrivals in any time slot is also bounded due to the bounded support assumption, the system state $X(t)$ takes on only finitely many values, though it will
be exponentially large. 

A scheduling policy $\pi$ has to choose, at each time $t$, possibly in a randomized way,
which packets to transmit at each node from the set of available packets, 
and over which links and at what powers.
The link choice allows routing to be optimized. The choice made at time $t$ will be denoted $U(t)$.
 
Since the probability distribution of the system state $X(t+1)$ at time $t+1$ depends only $X(t)$ and $U(t)$, the problem of maximizing the timely-throughput subject to node-capacity constraints~\eqref{shannon} is a CMDP, where a reward of $\beta_f$ is received when a packet of flow $f$ is delivered to its destination before its deadline expires:
\begin{align}
&\mbox{Maximize}_{\pi} \liminf_{T\to\infty}\frac{1}{T}\mathbb{E} \sum_{f}\sum_{t=1}^{T}\beta_f \delta_f(t) ,\notag\\ 
& \mbox{Subject to (\ref{shannon})}. \notag
%\tag*{(Constrained MDP)}
%\label{op}
\end{align}
Above, $\delta_f(t)$ is the number of packets of flow $f$ delivered in time to $d_f$ at time $t$. 

The above CMDP, parameterized by the vector $\beta:=\left(\beta_1,\ldots,\beta_F\right)$, is optimized by a stationary randomized policy ~\cite{altman1}. Since the numbers
of states and actions, are both finite, there is a finite set $\{\pi_1,\pi_2,\ldots,\pi_M\}$ of stationary randomized policies such that for each value of $\beta$, there is a policy that belongs to this set and solves the CMDP~\cite{altman1}. Let $\gamma_1,\gamma_2,\ldots,\gamma_M$ be the vectors of timely-throughputs associated with the policies $\pi_1,\pi_2,\ldots,\pi_M$. We then have the following characterization of $\boldsymbol{\Lambda}$:
\begin{lemma}\label{lemma1}
\begin{align*}
\boldsymbol{\Lambda} = \{ r: r=\sum_{i=1}^{M} \gamma_i c_i, c_i\geq 0,\sum_i c_i \leq 1    \}.
\end{align*}
\end{lemma}

Note that, though finite, using the above lemma is intractable for computation since the number of stationary Markov policies is exponentially large in the following parameter: Maximum possible number of packets in the network $\times$ Maximum path length of the flows $\times$ Maximum possible relative deadline. We therefore seek to design 
significantly lower complexity decentralized scheduling policies that achieve the region $\boldsymbol{\Lambda}$. 

Since we can restrict ourselves to stationary randomized policies, we can replace $\limsup$ and $\liminf$ by $\lim$:
\begin{lemma}
Maximizing the timely-throughput~\eqref{tp} subject to nodal average
power constraints~\eqref{shannon} can equivalently be posed as the following CMDP:
%\emph{Constrained Network-Wide Primal MDP}:
\begin{align}
&\Max_{\pi} \lim_{T\to\infty}\frac{1}{T}\mathbb{E} \sum_{f}\sum_{t=1}^{T}\beta_f d_f(t) \mbox{, subject to: } \notag\\ 
&\lim_{T\to\infty}\frac{1}{T}\mathbb{E} \sum_{t=1}^{T} \sum_{\ell:\ell=(i,\cdot)}E_\ell(t)   \leq P_i, \forall i\in \{1,2,\ldots,V\}.\label{op}
%\\
%\tag*{Primal MDP}
\end{align}
\end{lemma}

\section{The Packet-Level Decomposition of the Lagrangian of the Constrained Network-Wide MDP}\label{a1}
Defining $\lambda_i$ as the Lagrange multiplier associated with the power constraint on node $i$, and $\lambda := \left(\lambda_1,\lambda_2,\ldots,\lambda_{|V|}\right)$, we can write the Lagrangian for (\ref{op}) as,
\begin{align} \label{Lagr}
&\mathcal{L}(\pi,\lambda)= \lim_{T\to\infty}\frac{1}{T} \left\{ \mathbb{E} \sum_{f}\sum_{t=1}^T\beta_f \delta_f(t) \notag \right.\\
&\left. -\sum_{i} \lambda_{i}  \left(\mathbb{E} \sum_{t=1}^{T} \sum_{\ell:\ell=(i,\cdot)}E_\ell(t)   \right) \right\}+\sum_{i} \lambda_{i}P_i ,
%\tag*{Lagrangian}
\end{align}
where the expectation is w.r.t. the policy $\pi$ that is being used, the random packet transmission outcomes, and the randomness of the packet arrivals and relative deadlines, if the latter are random.

Denoting by $E_{\ell,f,n}(t)$ the amount of energy spent on transmitting the $n$-th packet of flow $f$ at time $t$ on link $\ell$, we have,
\begin{align*}
E_\ell(t) =  \sum_{f,n} E_{\ell,f,n}(t) .
\end{align*}  
The Lagrangian (\ref{Lagr}) reduces to,
\begin{align}
&\mathcal{L}(\pi,\lambda) = \sum_{i} \lambda_{i}P_i  \label{lang} \\
&+ \lim_{T\to\infty}\frac{1}{T}\sum_{f,n}\mathbb{E} \sum_{t=1}^T \left( \beta_f  d_f(t) -\sum_{i} \lambda_{i}  \sum_{\ell:\ell=i(i,\cdot)}E_{\ell,f,n}(t)  \right). \notag
%\tag*{Lagrangian}
\end{align}

This can be decoupled completely on a packet-by-packet basis for any fixed value of the vector $\lambda$, as follows.  

Let Packets$(f):=$ set of all packets of flow $f$. We will denote a packet by $\sigma$. Let Packets$(f,t):=$ subset of packets of flow $f$ that arrive before time $t$. Let $e(\sigma,i):=$ total energy consumed by packet $\sigma$ at node $i$, $\phi(\sigma):=$ flow that $\sigma$ belongs to, and $\delta(\sigma)$ be the random variable that assumes value one if packet $\sigma$ reaches its destination before its deadline and zero otherwise.  

Since the relative deadlines of packets are bounded, $ \mathcal{L}(\pi,\lambda)$ can be rewritten as a sum over packets:
\begin{align} 
 &\mathcal{L}(\pi,\lambda) = \sum_i \lambda_i P_i \label{Lagr2}\\
&+ \lim_{T\to\infty} \frac{1}{T}\mathbb{E} \sum_f \sum_{\sigma\in \mbox{Packets}(f,T)}	\left(\beta_f \delta(\sigma)-\sum_i \lambda_i e(\sigma,i)\right).			\notag
\end{align}   
The term corresponding to packet $\sigma$ of flow $f$,
\begin{align}\label{spt}
\mathbb{E} \left(\beta_f \delta(\sigma)-\sum_i \lambda_i e(\sigma,i)\right),
%\notag\\
%\tag*{Optimal Single-Packet Transportation Reward }
\end{align}
can be interpreted as follows.
The packet incurs a payment of $\lambda_i$ for using unit energy at node $i$, and accrues a reward of $\beta_f$ if
it reaches its destination before its deadline expires. 
This \emph{Optimal Single-Packet Transportation Problem}
is of very low complexity and is addressed further in Section \ref{spsf}.
Let $R(f)$ denote its optimal expected cost.

Due to the decomposition of (\ref{Lagr2}) over packets, we can optimize packet-by-packet. Hence we obtain,
\begin{align*}
\Max_{\pi} \mathcal{L}(\pi,\lambda) = \sum_f A_f R_f(\lambda) + \sum_i \lambda_i P_i,
\end{align*}
since $A_f$ is the arrival rate of packets of flow $f$.

%For each flow $f$, the term in the Lagrangian,
%\begin{align}\label{spt}
%\mathbb{E} \sum_{t=1}^T \left( \beta_f d_f(t) -\sum_{i} \lambda_{i}  \left[ \sum_{\ell:\ell=(i,\cdot)}E_{\ell,f,n}(t)\right] \right),
%%\notag\\
%%\tag*{Optimal Single-Packet Transportation Reward }
%\end{align}
%can be interpreted as follows. Each packet incurs a payment of $\lambda_i$ for using unit power at node $i$, and accrues a reward of $\beta_f$ (if of flow $f$) upon arriving at its destination. 

Therefore, for designing the policy $\pi$ for maximizing the Lagrangian, we can simply solve the Optimal Single-Packet Transportation
problem and let each packet make its own decision at each node on whether it
wants to be transmitted, and if so, at what power level.
No network state knowledge is needed by a packet to determine its optimal decision.
 Importantly, each packet's actions are independent of the actions chosen for all other packets in the network. It is very noteworthy that this results in a completely scheduling decentralized policy.

%\section{The Unit-Packet Unit-Flow Optimal MDP}\label{spsf}
\section{The Single-Packet Transportation Problem}\label{spsf}
%This gives rise to the following \emph{Optimal Single-Packet Transportation Problem}, addressed further in Section \ref{spsf}, involving transporting a single packet of flow $f$ from its source node $s_f$ to its destination $d_f$, so as to minimize the total cost for that packet. Every time it requests transmission by some node $i$ at a power level that consumes energy $E$, it has to pay $\lambda_iE$. As recompense, it receives a reward of $\beta_f$ if it successfully reaches its destination before its deadline. It has to make decisions at each time on whether to request transmission by the node at
%which it is located, and if so at what power level.

%Consider the single-packet cost expression~\eqref{spt}. 

%To simplify the discussion, let us consider the case when wireless fading is absent, i.e., the channel conditions are 
%static. The probability that a packet transmission over link $\ell$ at time $t$ is successful is given by $p_\ell(E)$. 
%Below, we fix attention on a packet of a particular flow $f$ and omit the subscript $f$. We relabel the nodes so that the source and destination nodes are labelled as $1$ and $V$ respectively.
The Optimal Single-Packet Transportation Problem is described as follows. A single packet of flow $f$ is generated at time $t=0$ at source node $s_f$, with state $(s_f, \tau_f)$, where $\tau_f$ is the time-to-deadline. At each time step thereafter, the time-to-deadline is decremented by one.
If it is not delivered to the destination node $d_f$ by time $t=\tau_f$, then it is discarded from the network. 
A reward of $\beta_f$ units is accrued if the packet reaches the destination node $d_f$ by time $\tau_f$.
%The age of the packet, $A(t)$, evolves as,
%\begin{align*}
%A(t+1) = A(t) + 1, \mbox{ if } A(t) < \tau_f,
%\end{align*}
%with the packet discarded if its age reaches $\tau_f$ units. 
A price $\lambda_i$ per unit energy has to be paid by the packet for transmission over an outgoing link at node $i$.  
%Thus,
%\begin{align}\label{cost:sp}
%& \mbox{One step cost at node $i$ at time } t = \notag \\
%& \begin{cases}
%\lambda_iE \mbox{ if transmission is attempted at time } t  \\
%\quad \quad \mbox{ at energy level } E \mbox{ over link } l=(i,j),\\
%0 \mbox{ otherwise},
%\end{cases}
%\end{align}
%while,
%\begin{align}\label{rew:sp}
%& \mbox{ Reward accrued at time $t$ at node $d_f$} = \notag \\
%& \begin{cases}
%\beta_f \mbox{ if delivered at destination node } d_f \mbox{ before deadline expiry},\\
%0 \mbox{ otherwise}.
%\end{cases}
%\end{align}
%The optimal single-packet transportation problem is to choose the control $U(t)$ so as to maximize the total reward
%\begin{align}\label{spp}
%\Max \mathbb{E}\left(\sum_{t=0}^{\tau_f} \beta_f \delta_f(t) -\sum_{i} \lambda_{i}  \left[ \sum_{\ell:\ell=(i,\cdot)}E_{\ell}(t) \right] \right),
%%\tag*{Optimal Single-Packet Transportation Problem }
%\end{align}
%where $\delta_f(t)$ assumes the value $1$ if the packet is delivered to node $d_f$ at time $t$, and is $0$ otherwise.

With the state of the packet described by the two tuple $\left(i,s\right)$, where $i$ is the node at which it is present, and $s$ is the time-till-deadline, we can use Dynamic Programming to solve for the value function $V^f$,
\begin{align}
V^f(i,s) &= \max \{ V^f(i,s-1), \notag \\
& \max_{j:(i,j)\in \mathcal{E},E}  \{-\lambda_i + p_{(i,j)}(E) V^f(j,s-1)  \notag \\
&  + \left(1-p_{(i, j)}(E) \right)V^f(i,\left(s-1\right)^{+}) \} \} , \label{bell} \\
V^f(d_f, s) &= \beta_f \mbox{ if } s \geq 0. \notag
\end{align}
Solving for the maximizer on the RHS yields the optimal action, i.e., whether to transmit or not, and if so, at what level,
for the packet of flow $f$ in the state $\left(i,s\right)$. 

It is important to note that this is a low complexity problem. Each packet's state only consists of the 
two tuple, (Node it is currently at, Time remaining to its deadline). This is simply a dynamic programming problem over a time horizon of $\Delta$, with $|V|\Delta$ states, in contrast to the exponentially large size,
$(V \Delta)^{F \Delta}$,
of the original network's state-space.
%, which we describe in the next section.

\section{The Dual of the Constrained Network-Wide Problem and Strong Duality}\label{pmdp}
The dual function is
\begin{align}\label{dual}
D(\lambda) = \Max_{\pi} \mathcal{L}(\pi,\lambda).
%\tag*{Dual Function}
\end{align}
Importantly, it can be obtained in a decentralized fashion for any price vector $\lambda$, 
due to the decomposition into a collection of Optimal Single-Packet Transportation problems coupled only through the node prices $\lambda_i$.

The Dual Problem is:
\begin{align}
\Max_{ \lambda \geq 0} D( \lambda ). \label{dualpro}
\end{align}
\begin{lemma}
There is no duality gap.
\end{lemma}
\begin{IEEEproof}
The CMDP~\eqref{op} can equivalently be posed as a linear program, in which the variables to be optimized are the \textit{occupation measures}~\cite{altman,ergodic1,ergodic2,altman1} induced by the policy $\pi$ on the joint state-action space. 
%Hence there is no duality gap.
Being a linear program, the duality gap 
%corresponding to the dual problem
%\begin{align} \label{dualpro}
%\max_{\lambda} D(\lambda)
%\end{align}
is zero. 
\end{IEEEproof}

%Let $\lambda^{\star}$ be the price vector that solves the Dual Problem~\eqref{dualpro}. It then follows from~\eqref{lagrangeeval}, that the policy $\pi(\lambda^{\star})$ solves the primal problem~\eqref{op}. We thus obtain
%the optimal solution as described in the following theorem: 
%
\section{Synthesizing Optimal Single-Packet Transportation Problems to Obtain the Network-Wide Optimal Policy} \label{solution}
 We now elaborate on how the optimal single-packet transport problem yields the overall network-wide optimal joint
 routing, scheduling and transmission policy.
 
 The key is to use randomization when packets are indifferent to being transmitted at two power levels.
 This arises when two different choices both attain the maximum of the RHS of the dynamic programming
 equation (\ref{bell}). In such cases, the action taken can be chosen randomly from one of the maximizers.
 Such randomization allows satisfaction of the power constraints with equality, or, to put it another way, it allows us to fully use up all
 the power that is available at a node if beneficial.

 \begin{theorem}
 Let $\lambda^\star \geq 0$ be a price vector. 
Denote by $\pi_f(\lambda^\star)$ an optimal randomized policy for packets of flow $f$, and by $\pi(\lambda^\star)$ the policy that implements $\pi_f(\lambda^\star)$ for each packet belonging to flow $f$.
% 
% Let $\pi\left(\lambda\right)$ be any optimal single-packet transportation policy, possibly randomized, that is optimal for the price vector $\lambda$, i.e., for $D(\lambda)$.
Suppose that, at every node $i$, either the average-power constraint (\ref{shannon}) is satisfied with equality by $\pi(\lambda^\star)$, or $\lambda^\star_i=0$.
Then $\pi(\lambda^\star)$ is optimal for CMDP, and $\lambda^\star$ is optimal for the Dual Problem.
 \begin{IEEEproof}
%For a vector of node prices $\lambda^\star$,  t
The dual function~\eqref{dual} is 
\begin{align}\label{lagrangeeval}
D(\lambda^\star) = \mathcal{L}(\pi(\lambda^\star),\lambda^\star).
\end{align}
The result simply follows from Complementary Slackness \cite{bertsekas} since the primal problem can be written
as a linear program over variables that are occupation measures.
% Let $R_f(\lambda)$ be the optimal reward of a packet of flow $f$ in the optimal single-packet transportation problem with nodal prices $\lambda$. 
% Consider $D(\lambda)$. It is a constrained MDP (\ref{dual}), 
%% :
%% \begin{align*}
%% \Max_{\pi} \mathcal{L}(\pi,\lambda), 
%% \end{align*}
% where
% \begin{align}\label{lang}
%&\mathcal{L}(\pi,\lambda)\notag\\
%&= \lim_{T\to\infty}\frac{1}{T}\sum_{f,n}\mathbb{E} \sum_{t=1}^T \left( \beta_f  d_f(t) -\sum_{i} \lambda_{i}  \sum_{\ell:\ell=i(i,\cdot)}E_{\ell,f,n}(t)  \right) \notag\\
%&\qquad+\sum_{i} \lambda_{i}P_i .
%%\tag*{Lagrangian}
%\end{align}
%
% \begin{align*}
% &\mathcal{L}(\pi,\lambda) :=\\
% & \lim_{T\to\infty} \frac{1}{T}\mathbb{E}^{\pi} \sum_{t=1}^{T}\left\{ [ \sum_f \beta_f d_f(t) ]+ \sum_i \lambda_i [ \sum_f E_f(t)-P_i  ]   \right\}    ,   
% \end{align*}
% $d_f(t)$ is the total number of packets of flow $f$ that are delivered to the destination in time, at time $t$, and $E_f(t)$ is the total energy used by packets of flow $f$ at node $i$ at time $t$. 
 \end{IEEEproof}
% Next we examine how to achieve tightness of all average-power constraints at all nodes.
\end{theorem}
%We consider the problem first assuming only a single route for each flow. Fix a price vector $\lambda$. Consider the flow $f$. Solve the Optimal Single-Packet Transportation Problem, and let 
%\begin{align*}
%p^f_{i,j}(s) = 
%\begin{cases}
%1 \mbox{ if transmitting from } i \mbox{ to } j \mbox{ is optimal}\\
%\mbox{  when the age of the packet is } s\\
%0 \mbox{ if transmission is not optimal}
%\end{cases}
%\end{align*}
%Note that if both transmission as well as no transmission are optimal, then we choose transmission above.
%Let $P^f(s)$ be the resulting matrix. Note that $P^f(s)$ is a sub-stochastic matrix. Let $d^f$ be the end-to-end deadline, and $\pi^f(0)= \left[0,\ldots 0 1 0 \ldots 0\right]$ denote the initial source of the flow. Then $\pi^f(0)P^f(0)P^f(1)\ldots P^f(d)e$ is the probability that the packet reaches the destination in time. Right multiplication by $e:=\left(1,1,\ldots,1\right)$ automatically takes the correct destination into account.

\begin{theorem}[Optimality of Decentralized Policy] \label{optimality-of-decentralized-policy}
The optimal policy for CMDP~\eqref{op} is fully decentralized. In order for any node $i$ to make a decision regarding a packet 
$\sigma$ present with it at any time $t$, the node only needs to know the flow $f$ that the packet belongs to, and the time-to-deadline of the packet. The optimal policy for node $i$ simply consists of implementing the policy $\pi_f(\lambda^{\star})$ for packets of flow $f$, where $\lambda^\star$ is the optimal price. 
\end{theorem}

We may observe the following key features of the solution. In order to solve the Primal Problem~\eqref{op} in its
original form, the network is required to make decisions based on the knowledge of the network state $X(t)$. The size of the state-space in which the network state $X(t)$ resides is exponential,
$(V \Delta)^{F \Delta}$, since there can be
$F \Delta$ packets in the network, with each being in one of $V \Delta$ states.
% a state in the following quantity
%and hence intractable: Number of packets ($\leq F \Delta$) $\times$ Deadline threshold bound $(\Delta)$ $\times$ Distance between nodes $(\leq |V|)$. 
Moreover, the optimal policy requires the entire network state information to be instantaneously known at each node at each slot. Indeed, one of the key reasons why optimal policies for communication networks (and other distributed systems) are generally intractable is that every decision requires instantaneous knowledge of the complete network state, which is something that cannot be obtained since the entire purpose of determining the optimal policy is to communicate information with deadlines. Thus, an approach based on directly solving the Constrained
Network-Wide MDP~\eqref{op} would have been computationally and implementationally futile.

These serious limitations have led us to instead formulate the 
Optimal Single-Packet Transportation Problem with the nodal transmission energy prices $\lambda_i^\star$.
%Dual Problem~\eqref{dualpro}. 
%The introduction of the nodal prices $\lambda_i^\star$ has the effect of reducing the 
This reduces the computational complexity from \emph{exponential} to \emph{linear}. Moreover the resulting solution can be implemented locally at each node. It is highly decentralized with no coupling between flows or nodes or even packets.

\section{Direct optimal solution of the network-wide problem} \label{direct-sol}
After establishing the above completely decentralized structure in Theorem~\ref{optimality-of-decentralized-policy},
we can directly obtain the network-wide optimal
policy by embedding the very low-dimensional single-packet transportation problems
of each flow into one flow-level problem, and then optimally allocating the power
at each node among all the flows.
We do this by considering the linear program involving ``state-action probabilities"
\cite{ros70b}.
In this approach we do not need to first explicitly solve for the optimal prices.

Let us consider a single packet of flow $f$. From Theorem~\ref{optimality-of-decentralized-policy} we can restrict attention to randomized Markov policies
where the packet is transmitted with a certain probability over a certain outgoing
link, or not transmitted at all, with the probabilities defending only on the state
$(i,s)$ of the packet. Note also that links are unreliable;
a packet transmitted over link $(i,j)$ reaches $j$ with probability $p_{ij}$.
Under a randomized Markov policy, the packet moves stochastically
through the network. We can delete a packet and remove it from the network as soon as $s$ hits 0. Let $\xi^f{(i,j,s)}$ denote the probability that the packet is transmitted
over link $(i,j)$ when its time-till-deadline is $s$, where we use the convention
that $\xi^f{(i,i,s)}$ is the probability that
it is not transmitted, and also define $p_{ii}=1$
correspondingly.
(Note that $s=0$ means that the this is the last allowed transmission of the packet, and
the packet will be deleted after this transmission).
The ``state-action probabilities" $\{ \xi^f{(i,j,s)} \}$
satisfy the constraints:
\begin{align}
&\sum_{j \in V, j \neq d_f} \xi^f{(j,i,s)} p_{j,i} +\sum_{m \in V} \quad\xi^f{(i,m,s)}(1-p_{i,m})  \notag \\
 &= \sum_{k \in V} \xi^f{(i,k,s-1)}  \quad \forall i \neq d_f, 
 1 \leq s \leq \tau_f , \label{direct-linear-constraints}
\end{align}
with the initial starting state $s_f$ captured by the equation
$
\sum_{j \in V} \xi^f{(s_f,j, \tau_f)} = 1 
$.
The probability that the packet reaches its destination $d_f$ before deadline expiry is
%\begin{align*}
$
\sum_{s=0}^{\tau_f} \sum_{i \in V} \xi^f{(i,d_f,s)} p_{i,d_f}
%\end{align*}
$.
If this policy is applied to all packets of flow $f$, then the average reward per unit-time
obtained by this policy is obtained by simply multiplying by the arrival rate $A_f$,
%\begin{align*}
$ \sum_{s=0}^{\tau_f} \sum_{i \in V} A_f \xi^f{(i,d_f,s)} p_{i,d_f}
%\end{align*}
$.
The energy consumed by a single packet at node $i$ is
%\begin{align*}
$
\sum_{s=0}^{\tau_f} \sum_{j \neq i} \xi^f{(i,j,s)} E
%\end{align*}
$.
The power consumption at node $i$ due to the packets of flow $f$ is
%\begin{align*}
$ \sum_{s=0}^{\tau_f} \sum_{j \neq i} A_f \xi^f{(i,j,s)} E
%\end{align*}
$.

Combining all the flows, we obtain the following \emph{Direct Linear Program} to
determine the optimal reward and the optimal Markov randomized policy:
\begin{align*}
& \Max \sum_{f \in F} \sum_{s=0}^{\tau_f} \sum_{i \in V} A_f \xi^f{(i,d_f,s)} p_{i,d_f} \\
& \mbox{Subject to:} \\
& \sum_{f \in F} \sum_{s=0}^{\tau_f} \sum_{j \neq i} A_f \xi^f{(i,j,s)} E \leq P_i 
\quad \forall i \in V, \\
%& \sum_{j \in V, j \neq d_f} \xi^f{(j,i,s)} p_{j,i} +\sum_{m \in V} \quad\xi^f{(i,m,s)}(1-p_{i,m})  \\
% &= \sum_{k \in V} \xi^f{(i,k,s-1)}  \quad \forall i \neq d_f, 
% 1 \leq s \leq \tau_f ,\\
& \sum_{j \in V} \xi^f{(s_f,j,0)} = 1 \quad  \forall f, \\
&  \xi^f{(i,j,s)} \geq 0 \quad \forall i \neq d_f,\mbox{ with }i,j \in V, f \in F, 0 \leq s \leq \tau_f, \\
&\mbox{ and equality constraints (\ref{direct-linear-constraints}) for all } f.
\end{align*}

This linear program directly determines the optimal power allocation over flows at each
node, and the optimal transportation policy for each packet. It also randomizes the
actions of all packets of a particular identically so that the power available is utilized optimally.

This is a low complexity linear program with only $|V|^2F \Delta$ variables
and $|V| + |V|F\Delta + F + |V|^2F \Delta$ constraints. This is a dramatic reduction of the
complexity, and is eminently tractable being a linear program.

\section{Optimality of a Threshold Policy} \label{threshold}
In fact, there is further structure that further reduces the complexity.
For simplicity we illustrate this when there is only one transmit power level that corresponds to a fixed energy usage $E$
for any transmission.
We show that each packet's decision is simply governed by a threshold on time-to-deadline.
\begin{theorem}[Threshold Policy]
For each flow $f$, and node $i$, there is a threshold $\tau_f(i)$, such that the optimal decision for a packet of flow $f$ at node $i$ with a time-to-deadline $s$ is to be transmitted/not transmitted according to
whether the time-to-deadline $s$ is strictly greater than/equal to or lesser than the threshold $\tau_f(i)$.
\end{theorem}
\begin{IEEEproof}
In a state where the decisions to transmit/not transmit are both optimal (i.e., the minimizer on the RHS of the Dynamic Programming equation (\ref{bell}) is not unique), we choose ``not to transmit," so that we thereby obtain an optimal policy that uniquely assigns an optimal action to each state. We will prove the following property (P) of this optimal policy, from which the theorem readily follows:
(P) If the optimal decision is to not transmit a packet at a node, then it is optimal to never again transmit that packet at that node.

The reason is that one can then simply define $\tau_f(i)$ as the maximum value of $s$ at which the decision to not transmit is the optimal action.
Now we prove property (P) by using stochastic coupling. Suppose that for a packet of flow $f$ at a node $i$ it is optimal to not transmit it at time-to-deadlines equal to $\sigma, \sigma-1, \ldots \sigma-k$, but it is optimal to transmit it
when its time-to-deadline is $\sigma-k-1$. Consider a packet, called Packet-1, that follows this optimal policy. It waits for $k$ slots at node $i$, and then gets transmitted when its time-to-deadline is $\sigma-k$. Now consider another packet, called Packet-2, that waits no time at node $i$, and is transmitted when its time-to-deadline is $\sigma$. We will couple the subsequent experiences of Packet-1 and Packet-2, i.e., whether a transmission at
a link is  failure or success, after that transmission.
Then, if Packet-1 reaches the destination $d$ in time, then so does Packet-2. Hence the reward accrued by Packet-2 is no less than the reward accrued by Packet-1, while its costs are the same. Hence the decision of Packet-2 to immediately get transmitted at time-to-deadline $\sigma$ is optimal. 
\end{IEEEproof}
However, one should not search for an optimal policy by trying to find the optimal thresholds. It is far better to search for optimal \emph{prices} since they are the \emph{same} for all packets of all flows, but the optimal threshold is only for a specific flow. That is, the prices result in the right trade-off between packets of different flows.
However, if one obtains a set of thresholds that is ``Person-by-Person" (or Nash) optimal for the flows, in the sense that each threshold is optimal for a particular flow when the thresholds of the other flows are fixed, then the entire set of thresholds may not be ``Team" Optimal. 
%That is why we search for the optimal prices, as in the next section.

\section{Determining the Optimal Prices}\label{sec:optpol}
%\subsection{The Dual of the Direct Linear Program}
Now we consider the problem of determining
the correct prices $\lambda^\star := \{ \lambda_i^\star: i \in V \}$ 
to be charged by the nodes. 
The key point to note is that price-determination can be done offline;
prices can be pre-computed and stored.

One method is to just obtain them simply as the sensitivities of the power constraints
in the Direct Linear Program. This requires a good model of the network and its reliabilities and the demands. However, one can also obtain them 
by ``tatonnement" over
a running system.
%The question that remains is how to choose 
Such price discovery is based on the Dual Function~\eqref{dual}.
First we discuss a hybrid of optimization and simulation, and subsequently a
purely learning approach.
%There are a variety of ways in which they can be computed, which we show below.

%Two approaches to determining $\lambda^\star$ suggest themselves; both can be conducted off-line.
%The optimal prices $\{ \lambda_i^\star \}$ can be pre-computed and stored, and then
%used to optimally operate the system.

\subsection{Repeated simulations}
Price discovery can be performed offline by repeated simulation. 
%For any fixed price vector $\lambda$ one can solve the dynamic programming equations for the optimal packet policy $\pi(\lambda)$. Subsequently, s
Since the Dual
Problem is convex, each node can use subgradient descent to
converge to the optimal price vector $\lambda^{\star}$. 
%The quantity $D( \lambda )$ can be estimated simply by simulation. Due to convexity, the 
The sub-gradient iteration is simply 
Walrasian tatonnement~\cite{walras1}, which drives the ``excess power consumption" at each node towards zero. Specifically, at the $n$-th iterate of the price vector, node $i$ chooses 
\begin{align*}
\lambda^{n+1}_i = \lambda^n_i + \epsilon [\mbox{Power consumed by node } i - P_i ].
\end{align*}
which simply amounts to a standard subgradient iteration \cite{boydbook} with a step-size $\epsilon >0$. 
%Once the prices are determined, during on-line operation each packet simply makes its own optimal decisions on each slot on whether it is beneficial to it to request the node that it is currently at to transmit it. This decision is independent of the decisions of other packets, and does not require any network state information.
For any fixed price vector $\lambda$ one can solve the dynamic programming equations for the optimal packet policy $\pi(\lambda)$.

\subsection{Employing on-line learning}
Instead of using simulation-based optimization to determine the optimal prices, one can determine both the optimal policy as well as the optimal policy contingent on that price, by using two time-scale stochastic approximation~\cite{robbins,borkarbook,kushner}. The faster time-scale stochastic approximation for the policy is
\begin{align}
& V_{n+1}(i,s) = 1_n(i,s) \left\{V_{n}(i,s) \left(1-a_n \right)\right. \notag \\
& \left. + a_n \max \{ V_n(i,\left(s-1\right)^{+}),X \} \right\} + \left(1-1_n(i,s)\right)V_{n}(i,s), \notag
\end{align}
where $X$ is the maximum, over all links $\ell = (i,j)$, of 
\begin{align}
 \lambda_{i,n} + p_{\ell}(E) V_n(j,\left(s-1\right)^{+}) + \left(1-p_{\ell}(E) \right)V_n(i,\left(s-1\right)^{+}). \notag
\end{align}
 In the above, $1_n(i,s)$ assumes the value $1$ if the packet-state at iteration $n$ is $(i,s)$. 
%In performing the above iterations, at iteration step $n$, a node needs to know the value function $V_n(\cdot)$ of its 
%neighboring nodes. 
 $\{ a_n \}$ is a positive sequence that satisfies $\sum_n a_n =\infty,\sum_n a^2_n<\infty$.  
% This is not a restriction since the iterations will be performed only at the commencement of network operation. Once the iterations converge, and the optimal policy is obtained, the policy can be implemented in a local fashion.

%Since the sub-gradient 
%%$\frac{\partial \mathcal{L}}{\partial \lambda_i}$ 
%evaluated at the policy $\pi(\lambda)$ is 
%equal to $ P_i-\bar{P}_i(\pi(\lambda))$, where $\bar{P}_i(\pi(\lambda))$ is the average-power utilization at node $i$ under the application of policy $\pi(\lambda)$, we 
We can use a slower time-scale stochastic approximation for the prices,
\begin{align}\label{pud}
\lambda_{n+1} = \lambda_n (1-b_n) + b_n (P-\bar{P}(\pi(\lambda_n))),
%\lambda_{n+1} = \lambda_n (1-b_n) + b_n (P-\bar{P}^{n}),
%\tag*{Price Update}
\end{align}    
where 
%$P= (P_1,P_2,\ldots,P_{|V|})$ 
$P$ is the vector consisting of nodal power bounds,
$\bar{P}_i(\pi(\lambda))$ is the average-power utilization at node $i$ under $\pi(\lambda)$, 
%$\bar{P}^{n}$ 
%the empirical average, 
and the sequence $b_n$ satisfies $\sum_n b_n = \infty,\sum_n b^2_n<\infty$, as well as $\sum_n (\frac{b(n)}{a(n)})^\gamma < \infty$, where $\gamma \geq 1$~\cite{borkarbook}. The iterations converge to the optimal prices $\lambda^\star$~\cite{borkarscale}.

When network parameters are not known, one can both solve for the optimal policy $\pi(\lambda^\star)$ as well as the optimal nodal prices $\lambda^{\star}$ in a decentralized manner. One way to achieve this task is to perform the Value Iterations using reinforcement learning for each price vector $\lambda$ until convergence, and then to update the price $\lambda$ using a gradient descent method. 

\section{Link-Capacity Constraints: The Truncation Policy for Near-Optimal Routing and Scheduling}\label{sec:linkcap}
The analysis so far has addressed 
%maximizing the network timely-throughput~\eqref{tp} subject to 
the nodal \emph{average}-power constraints~\eqref{shannon}.
%, in which the nodes are allowed to transmit multiple packets on their outgoing links. 
In this section, we impose more stringent peak link-capacity constraints on the number of concurrent packets that can be scheduled at any given time slot $t$.
$E_{\ell}(t)$, now re-defined
as the number of packets transmitted on link $\ell$ at time $t$, has to satisfy
\begin{align}\label{lcc}
E_{\ell}(t)\leq C_{\ell}, \forall \mbox{ links } \ell \in \mathcal{E},\mbox{ and } t=1,2,\ldots.
\end{align}
%If \eqref{lcc} holds at \emph{each} time $t$, then clearly it does so \emph{on average}. 
The more stringent problem that results is,
\begin{align}
&\mbox{Maximize}_{\pi} \lim_{T\to\infty}\frac{1}{T}\mathbb{E} \sum_{f}\sum_{t=1}^{T}\beta_f d_f(t), \notag\\ 
& \mbox{Subject to (\ref{lcc})}.
\label{op1}
%\tag*{Primal MDP}
\end{align}
We now proceed to construct a distributed policy with a provably close approximation to optimality. We begin with the policy that is optimal for the \emph{average} power constraint 
$$
\lim_{T \to \infty} \frac{1}{T} \mathbb{E} \sum_{t=1}^T E_\ell(t) \leq C_\ell.
$$
This is a distributed policy, as we have shown in the preceding sections, and is moreover tractable to compute. However it only ensures that the constraint~\eqref{lcc} is met on average, and \emph{not} at each time $t$. On the occasions that the number of packets that it prescribes for concurrent transmission does not exceed the constraint~\eqref{lcc}, we just transmit all the packets specified by that policy. However, on the occasions that it specifies an excessive number of transmissions exceeding the RHS of~\eqref{lcc}, we simply truncate the list of packets  in the manner to be specified below, and only transmit a total of $C_{\ell}$ of those packets. Clearly, this leads to a policy that does satisfy the constraint~\eqref{lcc} at each time instant. What we will show is that this policy is nearly optimal in a certain precise sense to be quantified below.

We first note the connection of our approach to that advocated by Whittle~\cite{whittle} for multiarmed bandits. Since there is no simple index policy~\cite{Git74} that is optimal when one is allowed to pull $n$ arms concurrently, if $n>1$, Whittle has suggested relaxing this constraint for \emph{each} time $t$ to a
constraint that the \emph{average} number of arms concurrently pulled is $n$. This relaxed problem has a tractable solution under an ``indexability" condition~\cite{Whittle2011Book}. Importantly, it is near-optimal when the number of arms available goes to infinity, with the proportion of arms of each type held constant~\cite{weiss}. Our approach can be regarded as an extension to multi-hop networks.

We first take care of one detail. The average-power constraints in~\eqref{op} are nodal, while the link-capacity constraints in~\eqref{op1} are link-dependent. To reconcile this, we consider the following version of the problem~\eqref{op} which involves average \textit{link-wise} power constraints,
\begin{align}\label{op2}
&\max_{\pi} \lim_{T\to\infty}\frac{1}{T}\mathbb{E} \sum_{f}\sum_{t=1}^{T}\beta_f d_f(t) \mbox{, subject to:} \notag\\ 
&\lim_{T\to\infty}\frac{1}{T}\mathbb{E} \sum_{t=1}^{T}E_{(i,j)}(t)  \leq C_{(i,j)}, \forall \mbox{links } (i,j)\in \mathcal{E}. %\tag*{Link-Wise Average Power Constraints}
\end{align}  
%Comparing the above with the problem~\eqref{op}, we notice that the bounds $P_i$ on average-power in~\eqref{op} are now replaced by the bounds $C_{(i,j)}$, with both the constraints being time-average constraints. It is easily verified that t
The optimal policy for the above problem can be obtained in exactly the same fashion as for the problem~\eqref{op}, except that now there are link-based prices $\lambda_{(i,j)}$, instead of nodal prices $\lambda_i$. 
%Thus, for example, the Optimal Single-Packet Transportation Problem now involves the following
%term in its dynamic programming recursion:
%\begin{align*}
%& \lambda_{(i,j)} + p_{(i,j)}(E) V_n(j,\left(s-1\right)^{+}) \\
%&\qquad \qquad \qquad + \left(1-p_{(i,j)}(E) \right)V_n(i,\left(s-1\right)^{+}).
%\end{align*}
%With the vector $\lambda^{\star}$ comprised of the optimal link prices $\lambda_{(i,j)}$ for all $(i,j)$, the optimal policy for the problem with link-wise average-power constraints is denoted by $\pi(\lambda^{\star})$ or simply $\pi^{\star}$. 

Now we define the truncation policy for the problem~\eqref{op1} which involves link-capacity constraints $\{C_{(i,j)}\}$: If the policy $\pi^{\star}$ specifies that node $i$ transmit more than $C_{(i,j)}$ packets at some time $t$, then the truncation policy can pick any $C_{(i,j)}$ of these packets and transmit them. Moreover, we eject from the network those packets which $\pi^{\star}$ dictated to be scheduled, but were not picked for transmission. (Discarding the packets is not strictly required, but it simplifies the discussion). Let us denote this modified policy by $\tilde{\pi}^{\star}$. It may be noted that under this policy, the evolution of the network is not independent across different packets, as was the case with $\pi^{\star}$.

For the policy $\pi^{\star}$, 
%the routing-scheduling decisions are packet-based, and taken only on the basis of age of the packet and its location in the network, 
let us denote by $p_f(\tau,\ell)$ the probability that under $\pi^\star$ a packet 
(of flow $f$) with age $\tau$ time-slots would be attempted on link $\ell$. Since the arrival rate of flow $f$ packets is $A_f$, then, on account of the fact that the policy $\pi^{\star}$ satisfies the average-power-constraints $C_{\ell}$ imposed by the network, we have
\begin{lemma}\label{l4}
\begin{align}
\sum_f \sum_{\tau=1}^{\Delta} A_f p_f(\tau,\ell) \leq C_{\ell}, \forall \ell \in \mathcal{E}.
\end{align}
\end{lemma}

Next, we determine the level of sub-optimality of $\tilde{\pi}^\star$.
\subsection{A Lower Bound on Performance of $\tilde{\pi}^\star$}
We will now obtain lower bounds on the performance of the policy $\tilde{\pi}^\star$. The following arguments are based on analysis of the evolutions of policies on an appropriately constructed probability space. Let us denote by $r_0$ the (average) reward earned by policy $\pi^{\star}$. First note that the reward collected by $\tilde{\pi}^\star$ (denoted by $r_1$) does not increase if it were to, instead of dropping a packet because of capacity constraint violation, schedule it as dictated by $\pi^{\star}$, but no reward is given to it if this packet is delivered to its destination node (denoted by $r_2$).
 However $r_2$ is more than the reward if now a penalty of $\beta_f$ units per-packet was imposed for scheduling a packet via utilizing ``excess capacity" at some link $l \in \mathcal{E}$, but it were given a reward in case this packet reaches the destination node (denoted by $r_3$). 
%Note that while calculating $r_3$, a reward is given upon delivery of a packet that was scheduled utilizing an excess bandwidth at any link in the network, but a penalty is imposed when the packet is scheduled using excess bandwidth at any of the links in the network. 
%But then, $r_3$ is simply the reward earned by the policy $\pi^{\star}$ in which a penalty of $\beta_f$ units is imposed on a packet if it is served using excess capacity \textit{at any node in the network during its stay in the network}.
$r_3$ is certainly more than the reward which $\pi^{\star}$ earns if it is penalized an amount equal to the sum of the excess bandwidths that its links utilize  (denoted by $r_4$) multiplied by $\beta_f$, since any individual packet may be scheduled multiple times by utilizing excess bandwidth. Thus, the difference $r_0-r_4$ is less than the sum of the excess bandwidths utilized by the links operating under the policy $\pi^{\star}$,
scaled by $\Max_f \beta_f$'s. Let $E_{f,\ell,\tau}(t)$ denote the number of packets of flow $f$ that have an age of $\tau$ time-slots, and are served on link $\ell$ at time $t$ under the policy $\pi^\star$.
 We thus have
 \begin{lemma}\label{l5}
 \begin{align}
  r_0-r_4 \leq \lim_{T\to \infty}\frac{1}{T}\mathbb{E} \sum_{t=1}^T \sum_{\ell \in \mathcal{E}} [ \sum_{f,\tau} 
 ( \Max_g \beta_g) \left( E_{f,\ell,\tau}(t)-C_\ell \right) ]^{+}.
 \end{align}
 \end{lemma}
\subsection{Asymptotic Optimality of $\tilde{\pi}^\star$ in the limit}
Next, we will scale the network parameters in two equivalent ways, and show that the policy $\tilde{\pi}^\star$ is asymptotically optimal. In the discussion below, $N$ is a scaling parameter. In the first approach to scaling, the link capacities $C_{\ell}$ and the mean arrival rates $A_f$ will be kept fixed, while the size of an individual packet will be scaled as $\frac{1}{N}$, with the arrivals being i.i.d. with binomial parameters $(N,A_{f} /N)$. The second equivalent formulation is to keep the size of packets fixed, while the link capacities for the $N$-th system are scaled as $NC_{\ell}$, with the arrivals being i.i.d. with binomial parameters $(N,A_f)$. In the rest of the discussion, we will confine ourselves to the former formulation; however a similar analysis can be performed for the latter case. 
%This scaling is akin to the fluid scaling technique utilized in the analysis of stability of queuing networks~\cite{bramson}. 

\begin{theorem}\label{th:2}
Consider the sequence of systems described in problem~\eqref{op2} parameterized by $N$, in which the arrivals for the $N$-th system are i.i.d. with binomial parameters $(N,A_{f} /N)$. The deviation from optimality of the $N$-th system in the sequence operating under the policy $\tilde{\pi}^\star$ is $O(\frac{1}{\sqrt{N}})$, and hence the policy $\tilde{\pi}^\star$ is asymptotically optimal for the joint routing-scheduling problem~\eqref{op1} under hard link-capacity constraints.
\end{theorem}
\begin{IEEEproof}
Below, $
MAD(X) := \mathbb{E}|X-\bar{X}|
$
is the mean absolute deviation of  $X$ with respect to its mean $\bar{X}$.

The deviation from optimality satisfies,
\begin{align*}
%\label{mad}
& r_0-r_4 \leq \lim_{T\to \infty}\frac{1}{T}\mathbb{E} \sum_{t=1}^T \sum_{\ell \in \mathcal{E}} \left( \sum_{f,\tau} \beta_f \left( E_{f,\ell,\tau}(t)-C_\ell \right)\right)^{+}\notag\\
&= \lim_{T\to \infty}\frac{1}{T}\mathbb{E} \sum_{t=1}^T \sum_{\ell \in \mathcal{E}} \left\{ \sum_{f,\tau} \beta_f \left( E_{f,\ell,\tau}(t)-\bar{E}_{f,l}(t)\right) \right. \notag\\
& \qquad \qquad \left. + \sum_f \beta_f \left( \bar{E}_{f,l}(t)-C_\ell \right) \right\}^+\notag
\end{align*}
\begin{align*}
&\leq \lim_{T\to \infty}\frac{1}{T}\mathbb{E} \sum_{t=1}^T \sum_{\ell \in \mathcal{E}} \left\{ \sum_{f,\tau} \beta_f \left( E_{f,\ell,\tau}(t)-\bar{E}_{f,l}(t)\right) \right\}^+\notag\\
&\leq \lim_{T\to \infty}\frac{1}{T}\mathbb{E} \sum_{t=1}^T \sum_{\ell \in \mathcal{E}} \sum_{f,\tau} \beta_f  \left\{ E_{f,\ell,\tau}(t)-\bar{E}_{f,l}(t)  \right\}^+\notag\\
&\leq \lim_{T\to \infty}\frac{1}{T}\mathbb{E} \sum_{t=1}^T \sum_{\ell \in \mathcal{E}} \sum_{f,\tau} \beta_f \left( \left| E_{f,\ell,\tau}(t)-\bar{E}_{f,l}(t) \right)\right|\notag\\
&= \lim_{T\to \infty} \frac{1}{T}\sum_{t=1}^T \sum_{\ell \in \mathcal{E}} \sum_{f,\tau} \beta_f MAD\left(E_{f,\ell,\tau}(t)\right)\notag\\
&= O\left(\frac{1}{\sqrt{N}}\right),
\end{align*}
where the 
%first inequality follows from Lemma~\ref{l5}, the second inequality from Lemma~\ref{l4}, and 
the last equality follows from~\cite{Moivre_closedform}.
%Denoting the reward earned by the optimal policy for the problem~\eqref{op1} by $r^{\star}$, we have, 
%\begin{align}
%r_0 > r^{\star}>r_1 > r_4.
%\end{align} 
%where the inequality $r_0 > r^{\star}$ follows from the fact that relaxing the link-capacity constraints~\eqref{op1} to link-wise time average constraints in~\eqref{op2} can only increase the network capacity, and $\pi^\star$ is optimal policy for the relaxed problem, while the inequality $r_0 >r_1 > r_4$ follows from Lemma~\ref{l5}. The proof is completed by noting that $r_0-r_4$ is $O(\sqrt{N})$.  
\end{IEEEproof}

\section{Near-Optimal Scheduling with Peak-Power Constraints}\label{sec:peakpow}
 We can similarly consider the problem of maximizing the network's timely-throughput~\eqref{tp} subject to 
 additional bounds on the \emph{peak} power that can be utilized by each node $i\in V$. 
%Note that the formulation in Section~\ref{rr} allowed each node $i$ to utilize arbitrarily high instantaneous power $\sum_{\ell:\ell=(i,j) }E_\ell(t)$ at time $t$. 
 The treatment is similar to that in Section~\ref{sec:linkcap}, in which we designed optimal policies under hard constraints on the \emph{number} of packets that can be transmitted at any given time $t$. 
 
 The problem is formally stated as,
 \begin{align}
 &\mbox{Maximize}_{\pi} \liminf_{T\to\infty}\frac{1}{T}\mathbb{E}\sum_{f}\sum_{t=1}^{T}\beta_f d_f(t)\mbox{, subject to } \label{peak:pow}\\ 
 &\limsup_{T\to\infty}\frac{1}{T}\mathbb{E}\sum_{t=1}^{T} \sum_{\ell:\ell=(i,j)}E_\ell(t) \leq P_i, \forall i\in \{1,\ldots,V\},\label{avgpow}\\
 &\mbox{ and }\sum_{\ell:\ell=(i,j)}E_\ell(t) \leq P^{\max}_i, \forall i\in \{1,\ldots,V\} \mbox{ and } t, \notag
 \end{align}
 where $P^{\max}_i$ (with value greater than $P_i$) is the peak-power constraint on node $i$.
 
 Denote the policy which maximizes the objective function~\eqref{peak:pow} under the average-power constraints~\eqref{avgpow} by $\pi^\star$. Now we modify it to a policy $\tilde{\pi}^\star$ described as follows: at each time $t$, each node $i\in V$ looks up the decision rule $\pi^\star$ and obtains the optimal power levels at which $\pi^\star$ would have carried out transmissions of the packets available with it. For this purpose, each node $i$ only needs to have knowledge of the age of packets present with it. Node $i$ then chooses a maximal subset of the packets present with it, such that the transmission power levels assigned to them by $\pi^\star$ sum to less than the bound $P^{\max}_i$. One way to choose such a set of packets and associated power levels is as follows: arrange the packets in decreasing order of the transmission power assigned by $\pi^\star$, and label them. Then $\tilde{\pi}^\star$ schedules the largest index packet such that the energy of all packets upto that index sum to less that $P_i^{\max}$. The asymptotic optimality of $\tilde{\pi}^\star$ follows as in Theorem~\ref{th:2}.
 \begin{theorem}
 Consider the sequence of networks operating under the policy $\tilde{\pi}^\star$, in which the arrivals for the $N$-th system are i.i.d. with binomial parameters $(N,A_{f} /N)$. The deviation from optimality of the $N$-th system in the sequence operating under the policy $\tilde{\pi}^\star$ is $O(\frac{1}{\sqrt{N}})$, and hence the policy $\tilde{\pi}$ is asymptotically optimal for the peak-power problem~\eqref{peak:pow}.
 \end{theorem}
 
 \section{Wireless Fading} \label{wirelessfading}
 Our model allows us to incorporate wireless fading.
%  by letting the link transmission success probabilities be a function of time $t$ and the energy $E$, i.e., $p_\ell(t,E)$. 
 We model the channel state as a finite-state Markov process $Y(t)$, with the link transmission success probabilities
 $p_\ell(Y(t),E)$ a function of the channel state $Y(t)$ and the transmit power level. As before, we assume that the probabilities
 are monotone decreasing in $E$. 
 
The network state is then described by a) the state of each packet, and b) the channel state $Y(t)$. The optimal policy can be determined along similar lines as before, by augmenting the system state with the channel state $Y(t)$. The optimal policy will be of the following form: the decision to be taken by a node $i$ at time $t$ will depend on the state of the packet and the channel state $Y(t)$. 
 
The above assumes that the channel condition is known to each transmitter. A simplification is possible if we assume that the process $Y(t)$ is i.i.d., which would eliminate the need for communicating $Y(t)$. Alternately it could
be deterministically time-varying. A common model which can be approximated is block fading~\cite{tvsn,rappaport}, under which the channel state need only be communicated periodically. 
 
 \section{Jointly Serving Real-Time and Non-Real-Time Flows}\label{jointnonreal}
 In the previous sections we have considered networks exclusively serving real-time flows for which the utility of a packet arriving after its deadline is zero. Often one is interested in networks that serve a mix of real and non-real-time flows~\cite{shakotaimix}. The system model can be easily extended.
%  to deal with the situation where a subset of the flows are real-time traffic, while the rest of the flows are non-real-time traffic. 
  To incorporate this case, we simply 
% modify the reward function, and the state evolution of the packets dicussed in Section~\ref{spsf}, for the packets 
 set the relative deadlines of the packets belonging to the non-real-time flows as $+ \infty$, so that they are never dropped.
 
% Thus the Optimal Single-Packet Transportation Problem would now be described as follows: a single packet is generated at the node $i$ at time $t=0$. Its age at time $t$, denoted $Y(t)$, increases by $1$ each time slot it is present in the network. Note that the state evolution differs from the case of deadline-constrained packets, since the packet is now not dropped if its age exceeds the threshold $d$. The packet exits the network upon arrival at the destination node. The cost and reward functions remain the same as in~\eqref{cost:sp},~\eqref{rew:sp}. All the analyses performed for the case of deadline-constrained flows, i.e., Value Iteration recursions, stochastic approximation based schemes, etc., carry over to the new setting.

\section{Illustrative Examples and Simulations} \label{examples}
%In this section, we consider some examples.

We first illustrate the amenability of the theory by explicitly hand-computing the
optimal distributed policy in two examples. In the second example, the deadlines
are slightly more relaxed than in the first example, and we show both how the prices change 
as a consequence, and
how the optimal policy reacts to this.

Subsequently, we consider a more complex example
and provide a comparative simulation illustrating the performance of the asymptotically
optimal policy for the case of link-capacity constraints, comparing it with 
well
known routing/scheduling policies
such as the Backpressure, Shortest Path, and Earliest Deadline First (EDF) policies.

\subsection{Two Illustrative Examples}
%We begin by illustrating the theory on an example, Example \ref{ex2}.
%We show how we 
%can explicitly determine the optimal solution.
%Subsequently we change the relative deadline and determine how the prices and optimal
%policy explicitly change as a result.
\begin{example}\label{ex2}
%% trim={<left> <lower> <right> <upper>}
%\begin{figure}[h]
%\includegraphics[trim = 70mm 80mm 20mm 40mm, clip, width=9cm]{figures/Fig-Ex-1.pdf}
%\caption{System considered in Example~\ref{ex2}.}
%\label{fig4}
%\end{figure}
Consider the system shown in Figure~\ref{fig4}.
\begin{figure}
\vspace{-0.7in}
\centering
\includegraphics[width=0.7\linewidth, angle=270]{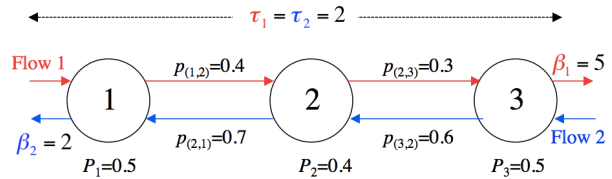}
\vspace{-0.65in}
\caption{System considered in Example~\ref{ex2}.}
\label{fig4}
\end{figure}
It consists of two flows traversing the Nodes 1, 2, and 3, in opposite directions.
Flow 1, with source node $s_1 =1$ and destination node $d_1 =3$, has an end-to-end deadline $\tau_1$ of 2 slots.
Flow 2, with source node $s_2=3$ and destination node $d_2=1$, also has an end-to-end deadline $\tau_2$ of 2 slots.
Packets cannot afford even one failure on any transmission if they are to reach their destinations in time,
since the relative deadlines for the flows are exactly equal to the total number of hops to be traversed.
One packet of each flow arrives in every time slot, so $A_1=A_2=1$.
Each packet transmission at any node is at 1 watt, so $E=1$ since all time slots are one second.
Nodes 1, 2 and 3, have average-power constraints $P_1=0.5, P_2=0.4$ and $P_3=0.5$ watts, respectively.
Links (1,2), (2,3), (2,1) and (3,2) have reliabilities of $p_{(1,2)}=0.4, p_{(2,3)}=0.3, p_{(2,1)}=0.7$ and $p_{(3,2)}=0.6$, respectively.
Denoting by $r_1$ and $r_2$ the timely-throughputs of Flows 1 and 2, we wish to maximize
$5r_1 + 2r_2$, i.e., packets of Flow 1 are 2.5 times more valuable than packets of Flow 2.
So $\beta_1=5$ and $\beta_2=2$.

The dynamic programming equations for the optimal single-packet transportation problem for Flow $1$ yield:
\begin{align*}
V^1(1,2) &= \Max \{0,-\lambda_1 + 0.4 V^1(2,1)\},\\
 V^1(2,1) &= \Max\{ 0,-\lambda_2 + (0.3)\cdot 5\}.
\end{align*}
So $V^1(2,1) = (1.5-\lambda_2 )^{+}$ and $V^1(1,2) = \left[\left(0.6-0.4\lambda_2\right)^+-\lambda_1\right]^+$.
Similarly, for Flow $2$,
$V^2(3,2) = \left[\left(0.84-0.6\lambda_2    \right)^+ - \lambda_3\right]^+$ and
$ V^2(2,1) = \left(1.4-\lambda_2\right)^+$.

Packets of Flow $1$ at Node $2$ are more valuable than packets of Flow $2$ at Node $2$,
since packets of Flow $1$ have expected reward of $(0.3)5=1.5$, while packets of Flow $2$ have expected reward of $(0.7)2=1.4$.
So we will push as many packets of Flow $1$ as possible to Node $2$.

In order for a packet of Flow 1 to choose to be transmitted at Node 2, however, the price $\lambda_2$ that it
pays needs to be less than the expected reward (0.3)5 that it can obtain in the future. Hence
\begin{align*}
\lambda_2 \leq 1.5.
\end{align*} 
Similarly, in order for a packet of Flow 1 to choose to be transmitted at Node 1, the total expected price it expects to
pay, $\lambda_1 + 0.4\lambda_2$ (since $\lambda_1$ is the price it pays at Node 1, and if it succeeds to reach Node 2,
which happens with probability 0.4, it then pays a price $\lambda_2$) must be less than the expected reward,
which is (0.4)(0.3)5. Hence,
\begin{align}
\lambda_1 + 0.4\lambda_2 \leq  0.6, \label{lambda1}
\end{align} 
But Flow $1$ can only push $(0.5)(0.4)=0.2$ of its packets to Node $2$. 
So there is spare capacity at Node $2$ that Flow $2$ can use. 
For Flow $2$ to use that we need $\lambda_2 \leq (0.7)2=1.4$. 
Now, Flow $2$ needs to utilize the spare capacity of $0.2$ left at Node $2$. 
So it needs to ensure a flow of $0.2$ reaches Node $2$. 
To do that it needs to transmit $\frac{1}{3}$ of the packets that arrive since $\frac{1}{3}(0.6)=0.2$. 
So it needs to randomize at Node $3$. 
By Complementary Slackness, this can only happen if packets at Node 3 are indifferent to being transmitted or not. So,
\begin{align*}
\lambda_3 +0.6\lambda_2 =(0.6)(0.7)(2) = 0.84.
\end{align*} 
Since we want to maximize $D(\lambda)$ we choose $\lambda_2=1.4,\lambda_3=0$, and, from (\ref{lambda1}), $\lambda_1$=0.04.

Therefore, we arrive at the following solution, where we denote by $\pi^f(i,s)$ the probability with which
a packet of flow $f$ is transmitted at Node $i$ when the time-to-deadline is $s$:
\begin{align*}
\lambda^\star& = (0.04,1.4,0).\\
\pi^1(1,2)&=0.5, \pi^1(2,1) = 1,\\
\pi^2(3,2) &= \frac{1}{3},\pi^2(2,1)=1.
\end{align*}

Now we verify that this policy is optimal.
$\lambda_2 =1.4$ implies $\pi^1(2,1)=1$ since $1.4\leq (0.3)5$. Now
$\lambda_1 +0.4\lambda_2 =0.6$ implies $\pi^1(1,2)=1$ and $\pi^1(2,1)=0$ are both optimal,
i.e., a packet is indifferent to them, and so one may randomize between them to satisfy the average-power constraint.
Similarly,
$\lambda_2=1.4$ implies that both decisions $\pi^2(2,1)=1$ and $\pi^2(2,1)=0$ are optimal. Also,
$\lambda_3 + 0.6\lambda_2 =0.84$ implies both $\pi^2(3,2)=1$ and $\pi^2(3,2)=0$ are both optimal.
So we can randomize the transmission of packets of Flow 2 in state $(3,2)$. 
The average-power usages are 0.5 watts at Node 1, 0.4 watts at Node 2, and $\frac{1}{3}$ watt at Node 3. 
The average-power constraints of $P_1=0.5$ and $P_2=0.4$ at Nodes $1$ and $2$, respectively,
are met with equality. The average-power constraint at Node $3$ is slack but $\lambda_3=0$.
So complementary slackness holds. 
Hence the policy is optimal. 
\end{example}
\begin{example}\label{ex3}
We now consider the same system as in Example \ref{ex2}, except that we relax the relative deadlines to
$\tau^1=\tau^2=3$, so that every packet can afford to have one hop that is retransmitted and still make it to its destination in time.
%The system is illustrated in Figure \ref{fig5}.

Consider a packet that has just arrived at Node 1.
It can either make it to its destination in two hops if both transmissions are successful the first
time they are attempted, or it can fail once at Node 1 and then be successful on subsequent transmissions at
Nodes 1 and 2, or it can succeed the first time at Node 1, fail once at Node 2, and then succeed at Node 2 on the second attempt.
If it does so reach its destination, it obtains a reward of 5. Hence taking these possibilities into account, if a packet of Flow 1
gets transmitted at every available opportunity, then the
Expected reward for a packet of Flow 1 at its first visit to Node $1 = [(0.4)(0.3)+ (0.6)(0.4)(0.3)+ (0.4)(0.7)(0.3)]5=1.38$.
Similarly, Expected reward for a packet of Flow 2 at its first visit to Node $3 = 1.428$,
Expected reward for a packet of Flow 1 at its second visit to Node $1 = 0.6$,
Expected reward for a packet of Flow 2 at its second visit to Node $3 = 0.84$,
Expected reward for a packet of Flow 1 at its first visit to Node $2 = 2.55$,
Expected reward for a packet of Flow 2 at its first visit to Node $2 = 1.82$,
Expected reward for a packet of Flow 1 at its second to Node $2 = 1.5$.
Expected reward for a packet of Flow 2 at its second visit to Node $2 = 1.4$.
% trim={<left> <lower> <right> <upper>}
%\begin{figure}[h]
%\includegraphics[trim = 70mm 80mm 20mm 40mm, clip, width=9cm]{figures/ex3.pdf}
%\caption{Network considered in Example~\ref{ex3}.}
%\label{fig5}
%\end{figure}

Packets of Flow $1$ are more valuable at Node $2$ than Flow $2$. So we want to maximize the throughput of packets of $1$ to Node $2$. If we transmit with probability $0.5$ on the first attempt at Node $1$ then all power is used up. The maximum power that can be consumed by packets of Flow 1 at Node 2 $= (0.5)(0.4)+(0.5)(0.4)(0.7)=0.34$ watts. So there is still $0.06$ watts left at Node $2$ that can be used by packets of Flow $2$. After arriving at Node 2 for the first time, a packet of Flow 2 can use a maximum power $1.3$ watts. 
So Flow 2 at Node 3 needs to make $\frac{0.06}{(1.3)(0.6)}$ attempts which amounts to 
randomization with probability $1/13$. 
In order to transmit a packet of Flow 2 on its second visit to Node 2, the price $\lambda_2$ cannot be any more than
the expected reward $(0.7)2=1.4$. 
So we could attempt some of the packets of Flow 2 that arrive at Node $3$, and transmit some packets
of Flow $2$ that arrive at Node $2$. 

With $\lambda_2=1.4$, $\lambda_1$ needs to satisfy 
$\lambda_1+(0.4)\lambda_2+(0.4)(0.7)\lambda_2 = \left[(0.4)(0.7)+(0.4)(0.7)(0.3)\right]5$, so
$\lambda_1 = 0.068$.
Similarly, $\lambda_3$ needs to satisfy,
$\lambda_3 + (0.6)\lambda_2+(0.6)(0.3)\lambda_2 = \left[(0.6)(0.7)+(0.6)(0.7)(0.3)\right]2$, which yields
$\lambda_3 = 0$.
The power constraint at Node $3$ is slack, but $\lambda_3=0$.
So the price vector is $\lambda=\left(0.068,1.4,0\right)$. The corresponding probabilities of transmission are
\begin{align*}
\pi^1(1,3) =0.5, \pi^1(1,2)=0,\\
\pi^1(2,2)=1, \pi^1(2,1)=1,\\
\pi^2(3,3) = 1/13,\pi^2(3,2)=0,\\
\pi^2(2,2)=1,\pi^2(2,1)=1.
\end{align*}
The optimal single-packet transportation dynamic programming equations yield:
\begin{align*}
V^1(1,3) &= \Max \left\{0,-0.068+(0.6)V^1(1,2)+0.4V^1(2,2) \right\} \\
&=0 \mbox{, so both choices are optimal,} \\
& \mbox{permitting randomization},\\
V^1(1,2) &= \Max \left\{0,-0.68+(0.6)(0)+0.4V^1(2,1)\right\}=0,\\
& \mbox{again, both choices are optimal,} \\
& \mbox{permitting randomization,} \\
V^1(2,2)&=\Max \left\{0,-1.4+(0.7)V^1(2,1)+(0.3)5\right\}=0.17,\\
V^1(2,1)&=\Max \left\{0,-1.4+(0.3)5 \right\}=0.1.
\end{align*}
In all of the below, both choices are again optimal,
\begin{align*}
V^2(3,3)=\Max \left\{0,0+(0.6)V^2(2,2)+0.4V^2(3,2) \right\}=0,\\
V^2(3,2)=\Max \left\{0,0+0.6V^2(2,1) \right\}=0,\\
V^2(2,2) = \Max \left\{0,-1.4+(0.3)V^2(2,1)+(0.7)2 \right\}=0,\\
V^2(2,1)=\Max\left\{0,-1.4+(0.7)2 \right\}=0.
\end{align*}
Note that the power consumptions are
\begin{align*}
P_1 & = (1)(0.5)=0.5 \mbox{, so it is tight,}\\
P_2 & =(0.5)(0.4)\left[1+(0.7)1\right]+\frac{1}{13}(0.6)\left[1+(0.3)1\right]=0.4 \mbox{, tight,}\\
P_3 & =\frac{1}{13}.
\end{align*}
The last constraint is loose, but then $\lambda_3=0$, and we still have complementary slackness. So the solution is optimal.
\end{example}

\subsection{Simulations}\label{simulations}

\begin{figure}
%\vspace{-0.5in}
\centering
\includegraphics[width=1\linewidth]{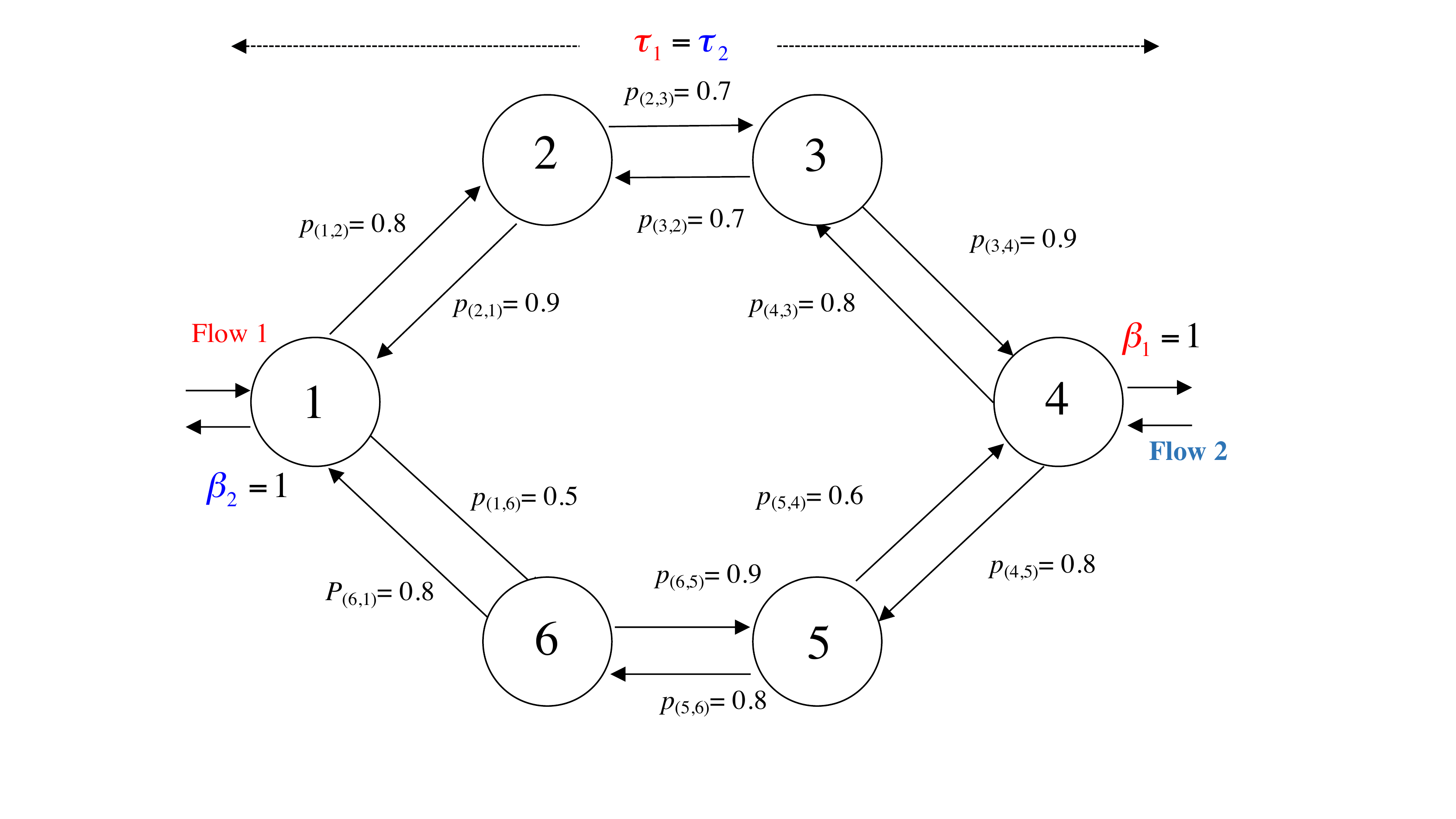}
\vspace{-0.4in}
\caption{Network with two source-destination pairs $(s_1=1,d_1=4)$ and $(s_2=4,d_2=1)$. Arrivals are deterministic with
rates $A_1=A_2=1$ per time-slot. Link capacities are $C_{(i,j)} \equiv 1$ packet/time-slot for all links $(i,j)$ shown.}
\label{fig6}
\end{figure}

\begin{figure}
\vspace{0.2in}
\centering
\includegraphics[trim = 60mm 50mm 20mm 20mm,width=0.4\linewidth, angle=270]{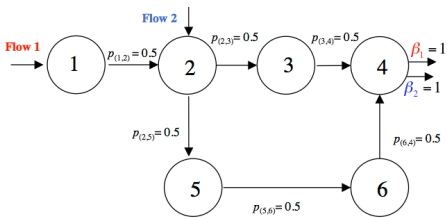}
\vspace{-0.35in}
\caption{Network with two source-destination pairs $(s_1=1,d_1=4),(s_2=2,d_2=4)$.  The arrivals are deterministic with rates $A_1=A_2=1$ per time-slot. Link capacities are $C_{(i,j)} \equiv 1$ packet/time-slot for all links $(i,j)$ shown.
}
\label{fig6.1}
\end{figure}

Now we consider the case of link-capacity constraints (or equivalently peak-power constraints).
We present a comparative simulation study of the asymptotically optimal policy with respect to the following two policies: a) Earliest Deadline First scheduling combined with Backpressure routing (EDF-BP), and b) Earliest Deadline First scheduling combined with Shortest Path routing (EDF-SP) that routes packets along the shortest path from source to destination with ties broken randomly.
We consider the systems shown in Figures~\ref{fig6} and~\ref{fig6.1}.  All link capacities are just 1, so
the asymptotically optimal policy is noteworthy for its
excellent performance seen below even in the very much non-asymptotic regime. 
% trim={<left> <lower> <right> <upper>}

%\begin{figure}[h]
%\includegraphics[trim = 60mm 50mm 20mm 20mm, clip, width=10cm]{figures/my-ex4.pdf}
%\caption{A system with two source-destination pairs $(s_1=1,d_1=4)$ and $(s_2=4,d_2=1)$. The arrivals are deterministic and fixed at $
%A_1=A_2=1$ per time-slot.
%The link capacities are set to $C_{(i,j)} \equiv 1$ packets/time-slot for all links $(i,j)$ shown as existing.}
%\label{fig6}
%\end{figure}
%
%
%
%\begin{figure}[h]
%\includegraphics[trim = 60mm 50mm 20mm 20mm, clip, width=10cm]{figures/ex4.pdf}
%\caption{A system with two source-destination pairs $(s_1=1,d_1=4)$ and $(s_2=4,d_2=1)$. The arrivals are deterministic and fixed at $
%A_1=A_2=1$ per time-slot.
%The link capacities are set to $C_{(i,j)} \equiv 1$ packets/time-slot for all links $(i,j)$ shown as existing.}
%\label{fig6}
%\end{figure}

We compare the performance of the asymptotically optimal policy $\tilde{\pi}^\star$ of Theorem~\ref{th:2},
with the following EDF-SP policy:
%that follows the Earliest Deadline First scheduling at each node, and routes packets along the Shortest Path from source to destination with ties broken randomly: 
%Under the EDF-SP policy, each node $i$ performs the following two operations in the order given below:
\begin{enumerate}
\item 
%It decides the link on which each packet available to it is scheduled based on the flow $f$ that the packet belongs to. 
The link $\ell =(i,j)$ chosen for scheduling packet transmissions for flow $f$ lies on the shortest path that connects the source and destination nodes of flow $f$.
\item  Thereafter, on each link $(i,j)$, it gives higher priority to packets 
%that are to be scheduled on it based on their deadlines, with higher
%priority given to packets 
having earlier deadlines. It then serves a maximum of $C_{(i,j)}$ packets in decreasing order of priority.   
\end{enumerate} 
We also compare the performance with the EDF-BP policy.
% which implements Earliest Deadline First scheduling at each node, and routes packets according to the Backpressure policy. 
 Under the EDF-BP policy, each node $i$ maintains queues for each flow $f$ and possible age $s$. 
Denoting by $Q_{i,f}(t,s)$ the queue length at node $i$ at time $t$, and by $Q_{f,i}(t) = \sum_s Q_{i,f}(t,s)$ the total number of packets of flow $f$ at node $i$ at time $t$, the policy functions as follows:
% Under the EDF-BP policy, each node $i$ performs the following operations:
\begin{enumerate}
\item For each outgoing link $\ell = (i,j)$, EDF-BP calculates the backlogs $Q_{f,i}(t)-Q_{f,j}(t)$ of flow $f$.
\item On each link $\ell =(i,j)$ it prioritizes packets on the basis of the backlogs associated with their flows. For packets belonging to the same flow, higher
priority is given to packets having earlier deadlines.
\item It then serves a maximum of $C_{(i,j)}$ highest priority packets from amongst the packets whose flows have a positive backlog  $Q_{f,i}(t)-Q_{f,j}(t)$.
\end{enumerate}
Both EDF-SP and EDF-BP eject packets that have crossed their deadlines.

Plots~\ref{fig7} and~\ref{fig9} show the comparative performances of the policies for the networks in Fig.~\ref{fig6} and~\ref{fig6.1} as the relative deadlines of the flows are varied. The performance of the asymptotically optimal policy is superior even in the non-asymptotic regime. Plots~\ref{fig8} and~\ref{fig10} show the comparative
performance as network capacities are increased. 
\begin{figure}
\vspace{-0.75in}
\centering
\includegraphics[width=0.7\linewidth, angle=270]{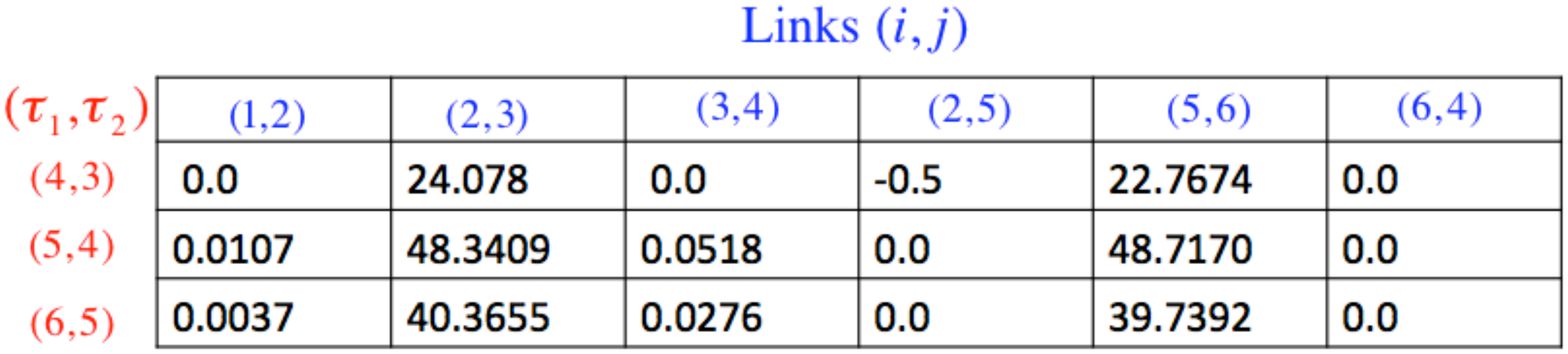}
\vspace{-0.75in}
\caption{Prices of links for the network in Fig.~\ref{fig6.1}.}
\label{sampleprices}
\end{figure}

% trim={<left> <lower> <right> <upper>}
Observe that for the network in Figure~\ref{fig6.1}, a
shortcoming of EDF-SP is that it is unable to utilize the path $1\to2\to 5\to 6 \to 4$, and therefore performs worse than EDF-BP.
%Thus, a shortcoming of the EDF-SP is that it will not be able to utilize all the existing source-destination paths in the network. 
Though it seems that in a general network the EDF-BP  should be able to utilize all source-destination paths, it will
neither be able to efficiently prioritize packets based on their age, nor discover which paths are more efficient at delivering packets within their deadlines. 
%This is one major drawback of the EDF-BP policy.  
\begin{figure}[h]
\includegraphics[width=9cm]{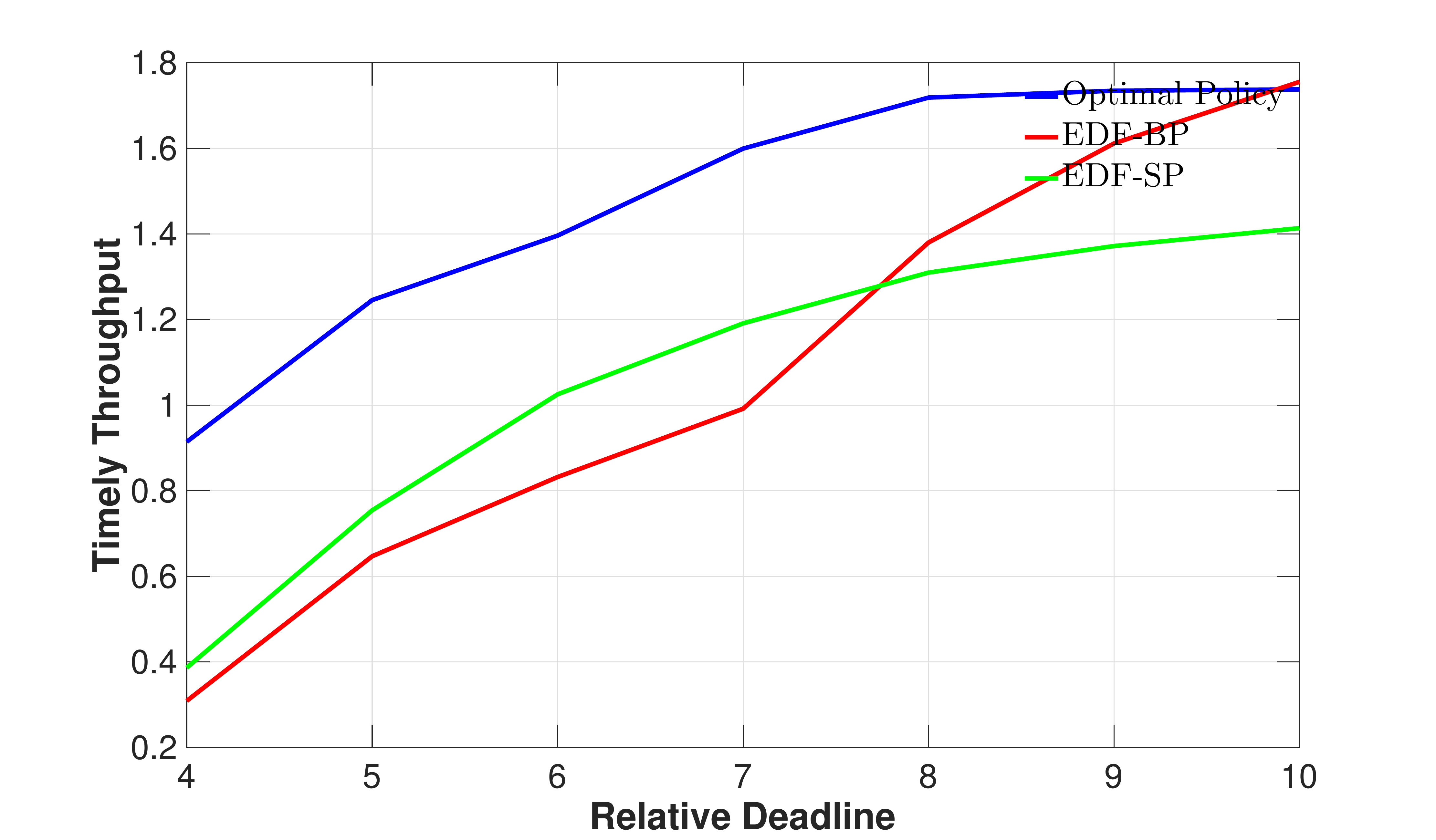}
\caption{Timely throughputs of the asymptotically optimal (labeled ``Optimal"), EDF-BP and EDF-SP policies for the network in Fig.~\ref{fig6} as the relative deadlines of both flows are increased.}
\label{fig7}
\end{figure}
\begin{figure}[h]
\includegraphics[width=9cm]{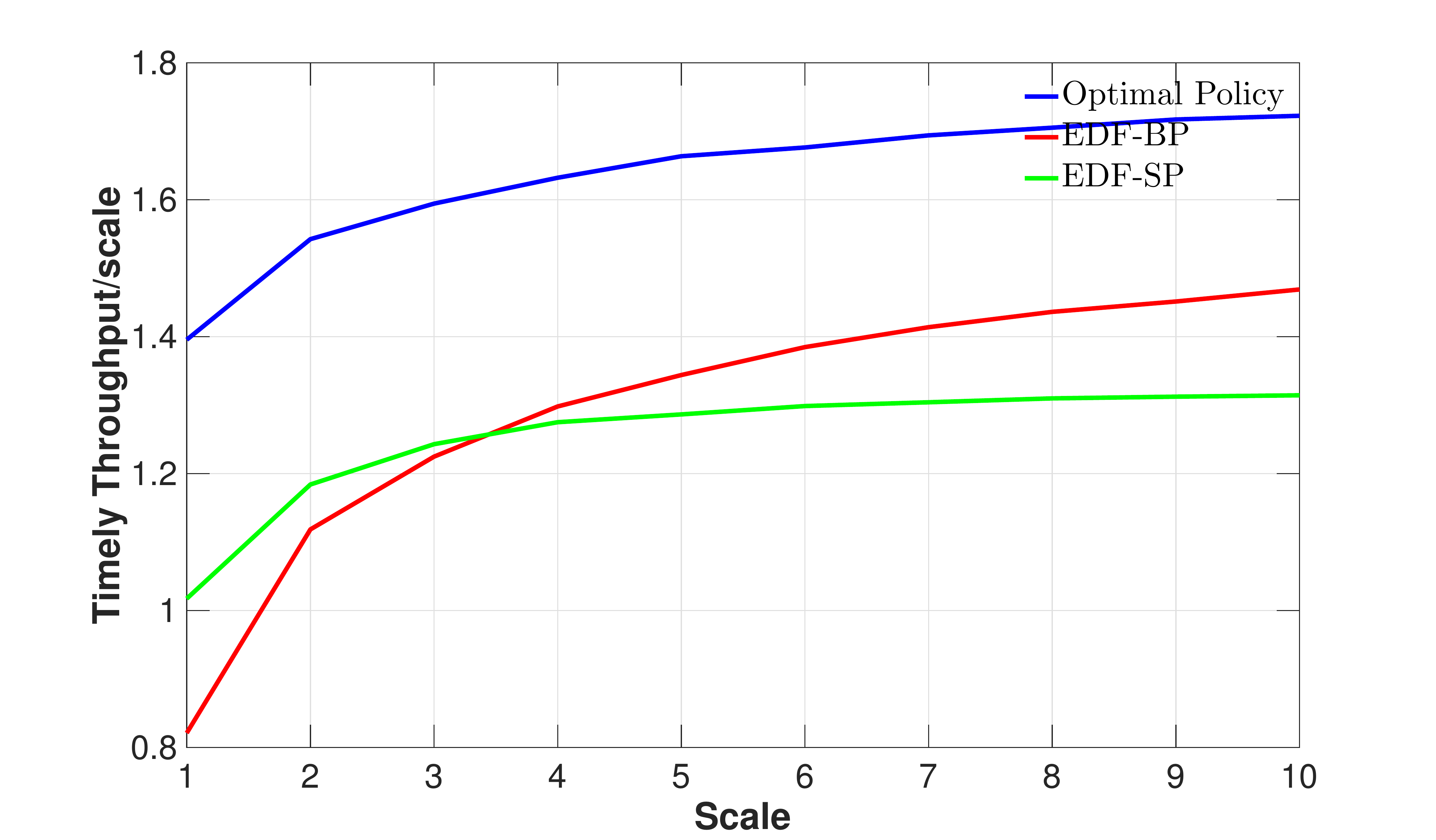}
\caption{Timely throughputs of the policies for the network in Fig.~\ref{fig6} as link capacities and arrival rate are scaled. The relative deadlines for both flows are set at $6$ time-slots.}
\label{fig8}
\end{figure}
\begin{figure}[h]
\includegraphics[width=9cm]{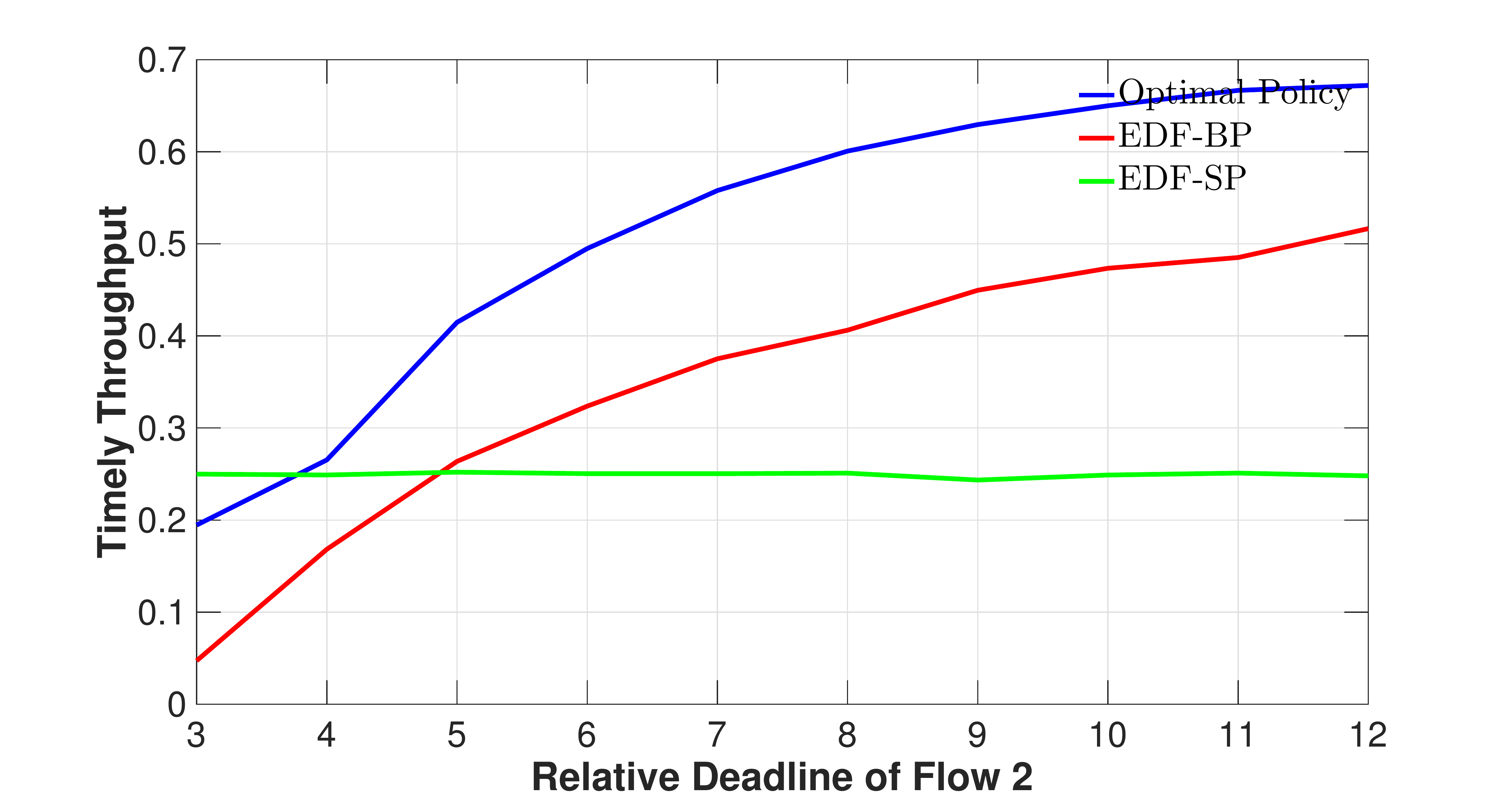}
\caption{Timely-throughputs for the network in Fig.~\ref{fig6.1} as the relative deadlines of flows are increased. Relative deadline of Flow 1 is one more than that of Flow 2.}
\label{fig9}
\end{figure}
\begin{figure}[h]
\includegraphics[width=9cm]{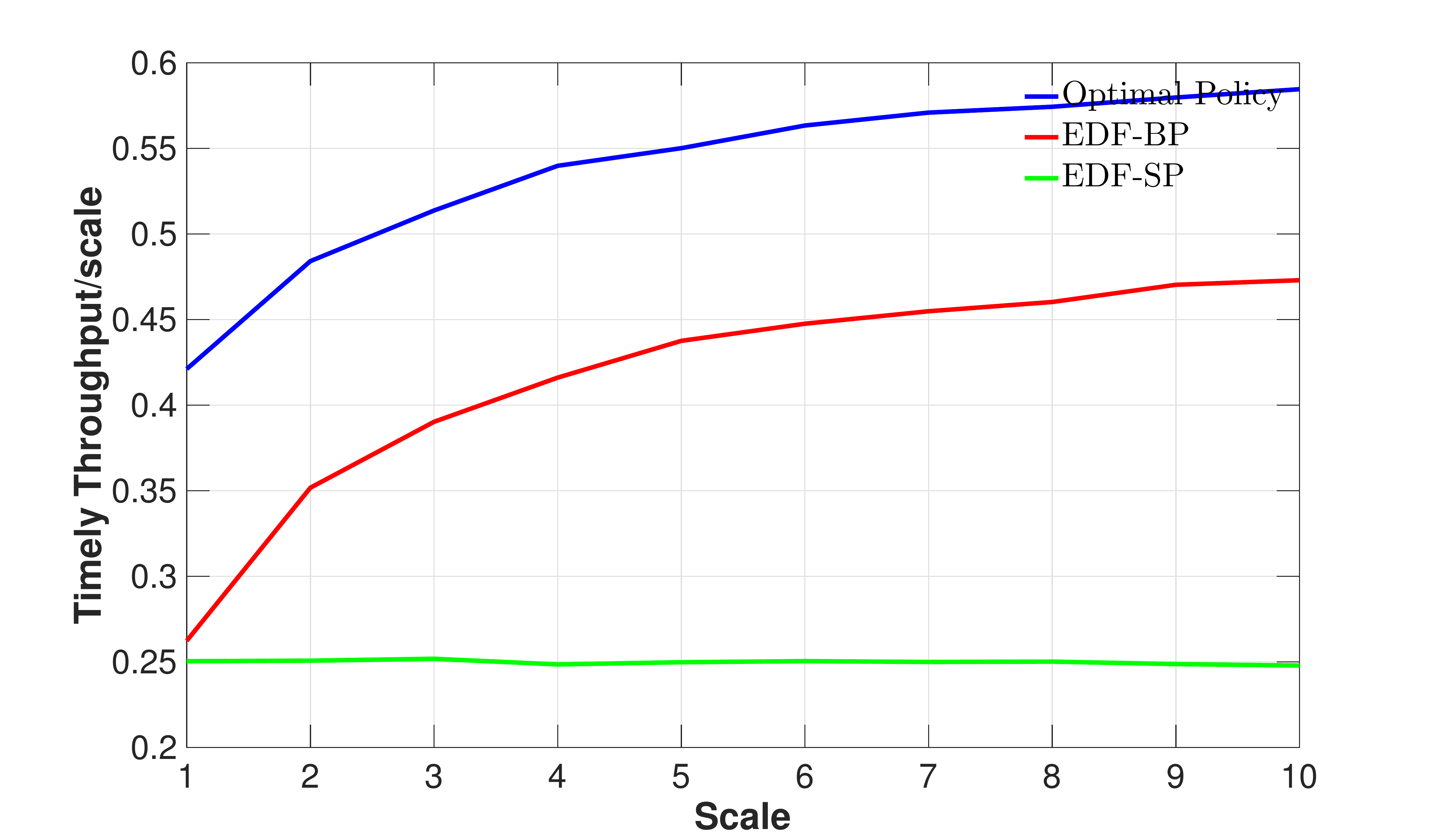}
\caption{Timely-throughputs for the network in Figure~\ref{fig6.1} as link capacities and arrival rates are scaled. Relative deadlines for Flows 1 and 2 are $6$ and $5$  respectively.}
\label{fig10}
\end{figure}

\section{Conclusions and Future Work} \label{conclusion}
 We have addressed the problem of designing optimal distributed policies that maximize the timely-throughput of multi-hop wireless networks with average nodal power constraints and unreliable links, in which data packets are useful only when they are delivered by their deadline. The key to our results is the observation that if the nodes are subject to average-power constraints, then the optimal solution is decoupled not only along nodes and flows, but also along packets within the same flow at a node. Each packet can be treated exclusively in terms of its time-to-deadline at a node. The decision to transmit a packet is governed by a ``transmission price" that the packet pays at each node, weighed against the reward that it collects at the destination if it reaches it before the deadline expires. 
  
%This approach extends to networks with peak-power constraints at each node. 
%%One simply relaxes them to average-power constraints, and truncates the resulting scheduling policy so that the bound is never exceeded. 
%A near-optimal distributed policy is designed.
%Its near-optimality can be precisely quantified in terms of a scaling of the network's flows.
 
 The above policies are highly decentralized; a node's decisions regarding a packet can be taken solely on the basis of its age and flow classification. The nodes need not share any information such as queue lengths, etc., amongst themselves in order to schedule packets. This approach is notable since obtaining optimal distributed policies for networks has long been considered an intractable problem.
% 
% The decentralized aspect of the derived policies is also surprising in view of the results in~\cite{tassi1}, where the Backpressure policy is shown to be throughput optimal, but the network nodes need to share the information of their queue lengths with neighboring nodes, in order to maximize the network throughput. In comparison, not only do our policies provide hard per-packet guarantees on the delays, but they are also highly decentralized. 
% 
Thus, our work fills two important gaps in the existing literature of policies for multi-hop networks a) hard per-packet end-to-end delay guarantees, b) optimal distributed policies.
 
The traditional approach to scheduling has been to consider the Lagrangian of the fluid model, and interpret the
queue lengths as prices. This addresses throughput optimality, but not delay, as one would expect from any fluid model-based analysis.
The key to our analysis consists of posing the problem of joint routing-
and scheduling packets under deadline constraints over a multi-hop network as an
intrinsically stochastic problem involving unreliabilities, and consider its Lagrangian and the Dual. 
This intrinsically captures variabilities in packet movement which critically
affect delays, and allows us to address the timely-throughput optimality of packets that meet
hard end-to-end deadlines.
The Lagrange multipliers associated with the average power or rate constraints are then
the prices paid by a packet to a node for transmitting its packet, rather than queue lengths. 
This yields a completely decentralized policy, in which decisions are taken by a packet based solely on its age and location in the network,
for which the accompanying dynamic programming equations are very tractable.
The overall solution is eminently tractable, being completely determined by a linear program with 
the number of variables equal to the product of the square of
the number of nodes, the number of flows and the maximum relative deadline,
rather than exponential in problem size.
 
% While dealing with the average-power constraints~\eqref{shannon}, we assumed that the nodes can transmit/receive packets simultaneously on multiple channels. This assumption has the following positive aspects: a) it enables the network to fully exploit the resource sharing techniques available to it, e.g., CDMA, OFDM, TDMA etc. b) it also allows the network to utilize the battery power available to it in a time-slot when it requires it the most, for example when a node has
% an excess of packets in a state in which it is desirable for them to be transmitted, c) this assumption vastly simplifies the construction of the optimal policy, which also turns out to be highly decentralized. The average-power constraint gives rise to a very simple model in which the constraints on number of links at the individual nodes’ disposal, and/or wireless power interference are ``smoothened/relaxed”. This smoothening has allowed us to treat the core issue of allocating the resources amongst the data packets in an optimal fashion. The model of average-power constraints is amenable to a treatment that shows us how a very simple decentralized policy can transfer data optimally if the network is willing to invest in resource sharing techniques mentioned above (e.g., TDMA, CDMA, OFDM etc.).
 
% Despite the usefulness of the average-power constraint in simplifying the analysis, it is not realistic to assume that a node can transmit any number of packets on a link,
% because in practice, there are only limited number of channels between any two nodes i and j in the network. 
We also consider the case of peak-power constraints at each node, which may be present in addition to, or as a replacement of,
average-power constraints.
%This means that the number of packets that can be attempted at any given time $t$, is strictly bounded by link-capacity $C_{(i,j)}$ , which motivates us to consider the model with link-capacity constraints. 
It is interesting that a minor modification of the optimal policy for the case
of average-power constraints is asymptotically optimal as the network capacity is scaled. 
%Furthermore, surprisingly, this policy for the case of link-capacity constraints is also highly decentralized. 
 
  This approach of dualizing the stochastic problem has broad ramifications, as has been explored in 
 \cite{SinKum15Video,SinKum15Storage} for problems such as video transmission and energy storage.
% [ATTENTION: List your other papers employing this approach, including the video problem, the free storage relaxation, etc]. 

This paper has considered only the case of unreliable links, which is of interest in networks
 consisting of microwave repeaters, networks with directed antennas, or even unreliable wireline links.
 The case of networks with contention for the medium is addressed in a companion paper.

\ifCLASSOPTIONcaptionsoff
  \newpage
\fi


\bibliographystyle{IEEEtran}
\begin{IEEEbiography}[{\includegraphics[width=1in,height=1.25in,clip,keepaspectratio]{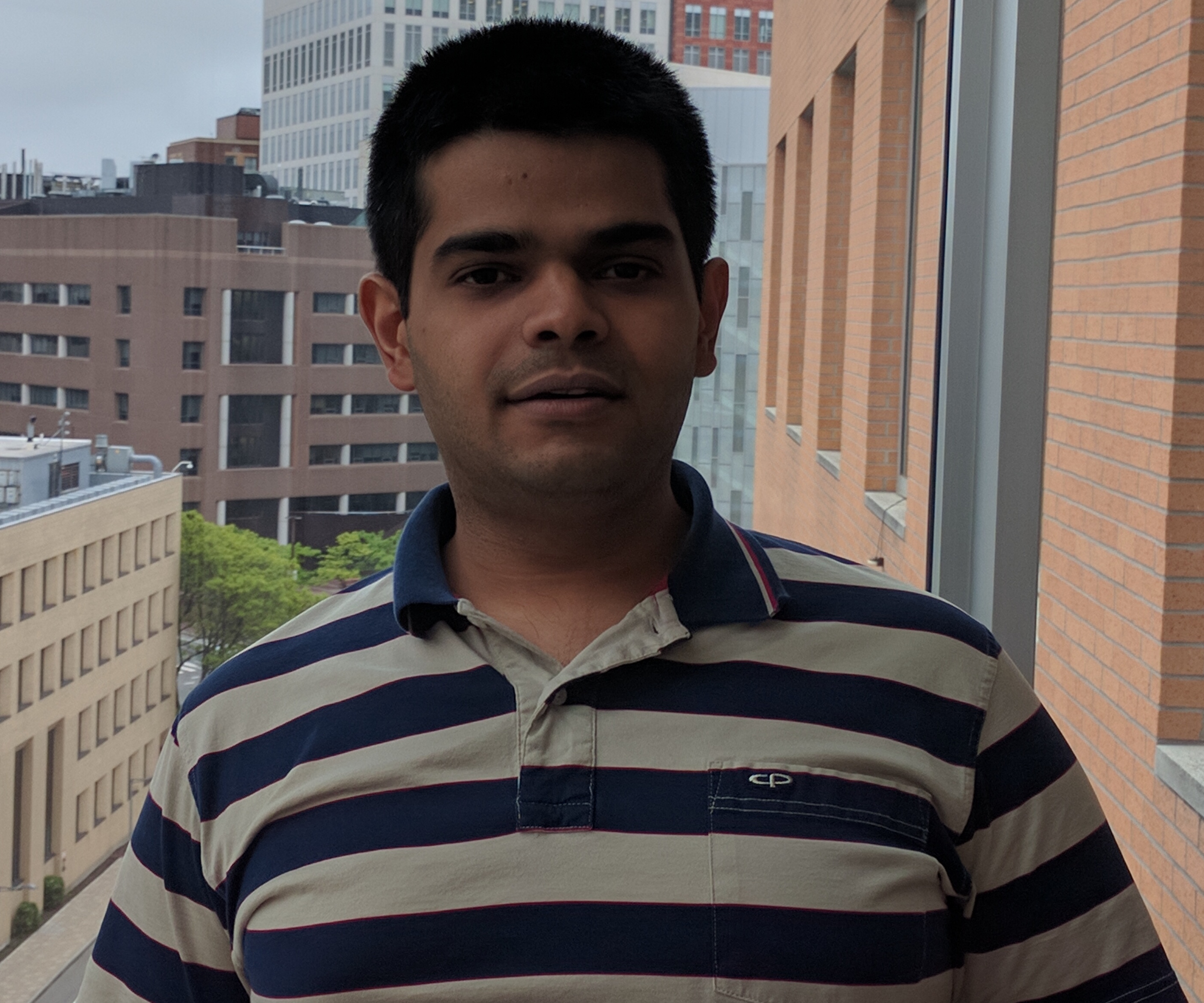}}]{Rahul Singh}
 received the B.E. degree in electrical engineering from
Indian Institute of Technology, Kanpur, India, in 2009, the M.Sc. degree in Electrical Engineering from University of Notre Dame, South Bend, IN, in 2011, and the Ph.D. degree in electrical and computer engineering from the Department of Electrical and Computer Engineering Texas A\&M University, College Station, TX, in 2015.

He is currently a Postdoctoral Associate at the Laboratory for Information Decision Systems (LIDS), Massachusetts Institute of Technology. His research interests include decentralized control of large-scale complex cyberphysical systems, operation of electricity markets with renewable energy, and scheduling of networks serving real time traffic.
%received the B.E. degree in Electrical Engineering from
%Indian Institute of Technology, Kanpur, India in 2009, the M.Sc. degree in Electrical Engineering from University of Notre Dame, South Bend, IN in 2005, and the Ph.D. degree in Electrical and Computer Engineering from the Department of Electrical and Computer Engineering Texas A\&M University, College Station, TX in 2015.
%
%He is currently a Postdoctoral Associate at the Laboratory for Information Decision Systems (LIDS), Massachusetts Institute of Technology. His research interests include decentralized control of large-scale complex cyberphysical systems, operation of electricity markets with renewable energy, and scheduling of networks serving real time traffic.
\end{IEEEbiography}
%\vspace{-0.4in}
\begin{IEEEbiography}[{\includegraphics[width=1in,height=1.25in,clip,keepaspectratio]{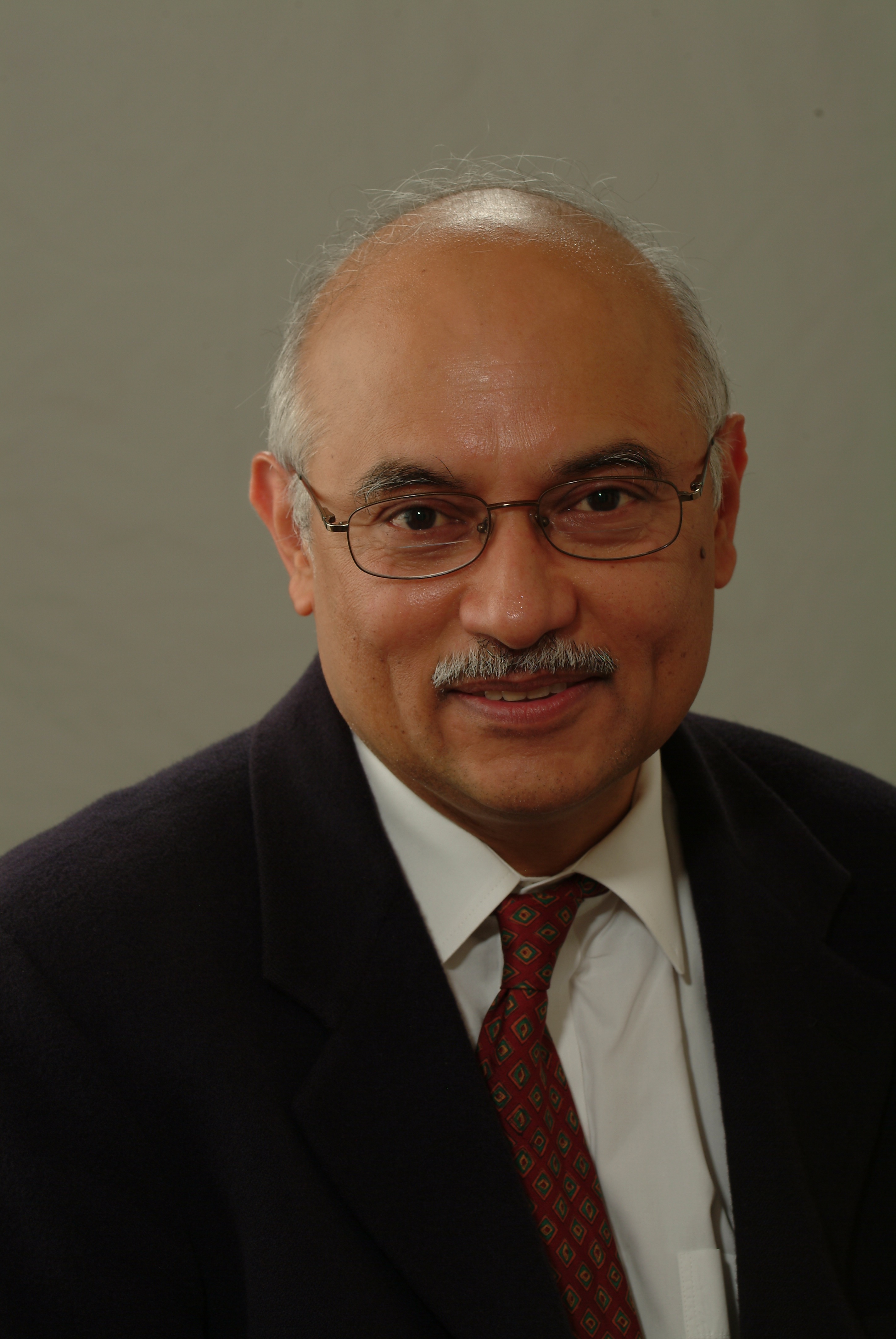}}]{P.~R.~Kumar} B. Tech. (IIT Madras, `73), 
D.Sc. (Washington University, St.~Louis, `77), was a faculty member at UMBC (1977-84) and Univ.~of Illinois, Urbana-Champaign (1985-2011). He is currently at Texas A\&M University. His current research is focused on stochastic systems, energy systems, wireless networks, security, automated transportation, and cyberphysical systems. 

He is a member of the US National Academy of Engineering and The World Academy of Sciences. He was awarded a Doctor Honoris Causa by ETH, Zurich. He as received the
IEEE Field Award for Control Systems, the Donald~P.~Eckman Award of the AACC,  Fred~W.~Ellersick Prize of the IEEE Communications Society, the Outstanding Contribution Award of ACM SIGMOBILE, the Infocom Achievement Award, and the
SIGMOBILE Test-of-Time Paper Award. He is a Fellow of IEEE and ACM Fellow. He was Leader of the Guest Chair Professor Group on Wireless Communication and Networking at Tsinghua University, is a D. J. Gandhi Distinguished Visiting Professor at IIT Bombay, and an Honorary Professor at IIT Hyderabad. He was awarded the Distinguished Alumnus Award from IIT Madras, the Alumni Achievement Award from Washington Univ., and the Daniel Drucker Eminent Faculty Award from the College of Engineering at the Univ.~of Illinois.
\end{IEEEbiography}


%\vfill


\end{document}